\documentclass{article}
\usepackage[utf8]{inputenc}
\usepackage[nottoc,numbib]{tocbibind}

\title{Quantum gravity in a laboratory?}
\author{\footnote{Authors listed in alphabetical order.} Nick Huggett, University of Illinois at Chicago\\ Niels Linnemann, University of Bremen / Rotman Institute of Philosophy\\ Mike D. Schneider, University of Illinois at Chicago}

\usepackage[margin=1in]{geometry}

\usepackage{natbib}
\usepackage{graphicx}
\usepackage{tikz}
\usepackage{amsmath, amssymb}
\usepackage{physics}

\usepackage[colorinlistoftodos]{todonotes}

\newcommand{\<}{\langle}
\renewcommand{\>}{\rangle}
\newcommand{\x}{\otimes}
\newcommand{\up}{\uparrow}
\newcommand{\dn}{\downarrow}

\usepackage{tikz}

\usepackage{url}
\usepackage{comment}

\begin{document} 

\maketitle

\tableofcontents

\section{Introduction}

The search for a fundamental theory of quantum gravity (QG) has prevailed for nearly a century. Despite much progress on the theoretical side --- including whole avenues of research toward how to develop such a theory (e.g. loop quantum gravity or string theory), little has been claimed on the empirical side. According to standard lore, this is entirely unsurprising given the Planck energy scale compared to the energy scales probed in high-energy particle physics; \cite{zimmermann2018future} for instance pointedly illustrates the remoteness of the Planck energy scale for our usual collider technologies based on acceleration of charged particles in electric and magnetic fields as follows:
\begin{quote}
An ultimate limit on electromagnetic acceleration may be set by the Sauter-Schwinger critical field, above which the QED vacuum breaks down. ... Assuming these fields, the Planck scale of $10^{28}$eV can be reached by a circular or linear collider with a size of about $10^{10}$m, or about a tenth of the distance between earth and sun, for either type of collider (!). (p. 36-37, exclamation mark in original)
\end{quote}
But maybe the emphasis on collider engineering limitations fails to communicate just how remote is the Planck energy scale from our more familiar physics. Here, an astrophysical benchmark may be further help: the phenomenon of Hawking radiation, by whose detection we would like to corroborate the formal apparatus of quantum field theory in curved spacetime (merely \emph{on the way to} a theory of quantum gravity apt for the Planck energy scale) is so weak that ``Trying to detect astrophysical Hawking radiation in the night’s sky is thus like trying to see the heat from an ice cube against the background of an exploding nuclear bomb" (\cite{thebault2016can}, p. 4).

What is the empirically-minded QG researcher to do? Despair is never an option, but maybe desperation is: one could search for an evidential or confirmatory framework that leaves room for \emph{non}-empirical forms of support for developments on the theoretical side of quantum gravity research. From this perspective, the framework for non-empirical theory confirmation developed by \cite{dawid2013string} may be attractive; indeed, something like this desperation, with regards to the status of a string theory approach to quantum gravity research, would seem to have been the motivation for the string theorists whose methodology Dawid describes. 

Alternatively, one could follow the sub-community of (self-styled) \emph{QG phenom\-enologists}: those empirically-minded quantum gravity researchers who have not stopped working on finding empirical signatures of QG, despite acknowledging the naive estimate of the difficulties as that of, say, Zimmermann quoted above. This is the tack we intend to take here.

One strategy in quantum gravity phenomenology is to look for QG effects within traces of high-energy astro-particle phenomenology in the early universe (famously motivated under the guise of the universe being, in light of standard model cosmology, the `poor man's accelerator'). Another strategy is to systematically search for effects that `cascade' from high energies to low energies, such as in many cases of Lorentz invariance violation (in either the astrophysical-cosmological arena or in more controlled experimental settings). On this strategy, one accepts that the relevant energy scale is the Planck scale (and, as such, is extraordinarily remote), but one contemplates rejecting an otherwise tacitly assumed fact of decoupling.\footnote{In particle physics, there is a `decoupling theorem' that relies on the framework of quantum field theory. Just so: this strategy in quantum gravity phenomenology explores QG effects beyond field theory (in the relevant sense).} A third strategy though, which will become our focus here, has only recently become relevant. It begins by noting that the Planck mass, rather than the Planck energy, might better serve as the quantity of interest in probing the quantum nature of gravity. As \cite{christodoulou2019possibility} write:
\begin{quote}
Puzzling is the fact that --- unlike Planck length and Planck energy --- $m_{\text{Planck}}$ falls within a very reachable physical domain: micrograms. It has long been hard to see what sort of quantum gravity effect can happen at the scale of the weight of a human hair. (p. 65)
\end{quote} 
It has long been hard, but perhaps (evidently) it will not be so hard any longer. And so, QG effects might indeed be in reach of tabletop experiments. Or at least, this is what recent claims amount to, in the emerging experimental research program in QG phenomenology which we will dub \emph{tabletop quantum gravity}.\footnote{In fact, the experiments will likely be rather too large literally to fit on a normal table: a large table, to be sure --- but at least not solar-system sized!}

It is important to note that thus restricting attention to the weak-field, Newtonian regime produces a subtle but significant change in the question of empirical access: one is no longer probing fundamental QG, only the low energy physics that different fundamental theories likely have in common. The question that faces us then is how might we read distinctively quantum traces of gravitational physics in such experiments? Answering this question seems to be key in making sense of the nascent tabletop quantum gravity research program. And in fact, only recently has it become clear that there is significant dissent among those physicists interested in quantum gravity phenomenology over the answer to this question. A new call to interpret hypothesized results in a proposed class of tabletop quantum gravity experiments, which \cite{bose2017spin} and \cite{marletto2017witnessing} have each independently noted may soon be viable, brings the question to the fore. 

In this new class of \emph{gravitational induced entanglement} (GIE) experiments (sometimes, Bose-Marleto-Vedral experiments) one employs spatially separated pairs of `gravcats'\footnote{The name `gravcats' goes back to \cite{anastopoulos2014problems}.}, or gravitationally coupled Schrödinger cats (i.e. superposition states of macroscopic, uncharged massive bodies), as the relevant quantum matter probes. Within these experiments, the hypothesized role for the underlying \emph{quantum nature} of Newtonian gravity is to mediate entanglement between the two gravcats in the pair. The proposal, then, is that if such experiments indeed produce the predicted gravcat entanglement, then this outcome would provide the first ever laboratory \emph{witnesses} of the quantum nature of gravity. And while this achievement would not amount to a direct observation of QG (and especially not to a direct observation that would distinguish between various current approaches to developing a fundamental theory of QG), it would still seem to be an enormous advance. Yet, even the nature of this achievement in terms of a first tabletop quantum gravity witness is questioned by some in the community. 

The stage thus set, three philosophers of physics came together, hoping to clear up for themselves a puzzle. How could it be that this specific, newly proposed class of experiments in tabletop quantum gravity could be a locus of dispute when all those involved in the dispute would seem to agree on their expected outcomes? And what do those results have to do with the supposed underlying \emph{quantum} nature of gravity, anyway? The manuscript compiled here is our best attempt, as those three philosophers, to provide a satisfying, unified answer to both of these interrelated questions. It is an answer that the three of us are, finally (after considerable friendly disagreement between us), content each to call our own. 

The preceding paragraph is as much context for the content of the project presented here as it is an apology for the presentation's length and discursive approach. That which might have been a series of idiosyncratic articles written by each of us in turn, arguing back and forth through a thicket of distractions, has instead been absorbed into the singular, long-form manuscript here, for which we may now provide a succinct roadmap. Following this Introduction, we offer a theoretical prelude ($\S$\ref{sec:prelude}) on so-called \emph{semi-classical gravity} --- a bit of theoretical architecture relevant for research in quantum gravity phenomenology, but whose conceptual status within our current best physics is equivocal. We distinguish three views of semi-classical gravity, or (in fact) two different ways of denying that semi-classical gravity is itself to be understood as a candidate for future fundamental theory in the discipline, given our current best physics today. In parallel with these two denials, we then provide in $\S$\ref{sec:experimentalprelude} an experimental prelude, noting two experiments in the early history of tabletop quantum gravity that are by now unavoidable in any conversation about quantum matter probes in a Newtonian gravitational context. Crucially, we will explain how these two experiments are importantly distinct from each other --- particularly in the kinds of conclusions drawn from their successful execution. This observation occasions our identifying two traditions of experimental testing that will become relevant in our assessment of the GIE experiments, beginning in the section thereafter: on one tradition, the goal is to witness the quantum nature of gravity; on the other, one rather is interested in control of (or access to) the same.

With these preludes in place, we turn then to the GIE experiments, and our analysis spans the remaining four sections before the conclusion. After a preliminary naive rehearsal of the GIE experiments in $\S$\ref{secGIE}, including a discussion of how they indeed would rule out semi-classical gravity as a candidate fundamental theory), we then turn to comment on the central question at stake when the experiments are considered in the tradition of witnessing ($\S$\ref{sec:paradigms}): to what extent may we take the experiments as capable of witnessing a quantum nature of the gravitational state? We ultimately argue that one's answer to this question very much hinges on a choice of modeling `paradigm' for the GIE experiments (a term we use carefully), even given agreement about our current best physics. In particular, we develop what we have come to understand are the two major such paradigms in play in the relevant literature: what we call the `Newtonian model' paradigm and `tripartite models' paradigm, respectively. Only according to the latter does gravcat entanglement actually witness the quantum nature of gravity. We then consider in $\S$\ref{ssec:Belenchia} a further question of whether one may conclude that what is witnessed in the GIE experiments, within the tripartite paradigm, are (coherent states of) \emph{gravitons} in a `low-energy' theory of QG. Finally, in $\S$\ref{secMakingQuantum}, we shift gears to offer the suggestion that the GIE experiments may at least as well be conceived in terms of their standing in the tradition of controlling and accessing --- rather than witnessing --- the quantum nature of gravity. And then we conclude. 

Taking a step back, our first goal is thus to inspect the claim that a GIE experiment could amount to a tabletop quantum gravity witness, in light of the dissent found within the physics community. We will find, on disentangling the various threads in the literature, that there is meaningful ambiguity as to whether the predicted outcomes of these experiments, if successful, would indeed provide such a witness. In particular, one's assessment depends on how one chooses to model the experimental setup, while our current best physics provides justifications for two distinct choices. However, our second goal is to provide a view of the GIE experiments that we believe adequately captures their punchline, as a matter at the frontiers of experimentation in tabletop quantum gravity, and which critically does not depend on choice of paradigm. Our hope in writing this manuscript is thus also to clarify that the successful completion of a tabletop quantum gravity experiment would be an enormous achievement for the experimental research program, regardless of further disagreements regarding the matter of witnessing.

\section{Theoretical prelude: `semi-classical gravity'}\label{sec:prelude}

The problem of QG is generally understood as a need to unify two elements of our current corpus of fundamental physics: on the one hand a classical and geometrized theory of gravity, general relativity (GR), which recovers Newtonian gravitation at low velocities; while on the other hand, a quantum theory of (special) relativistic matter, the standard model of particle physics, which recovers non-relativistic quantum mechanics (NRQM) in a markedly different low-velocity limit.\footnote{The difference being that, in the latter case, we also need to recover at those low velocities a theory of finite degrees of freedom from a theory of infinite degrees of freedom.} But matter explicitly appears in GR as classical, contrary to our simultaneous embrace of a quantum field theory (QFT) description of matter, in the form of the standard model of particle physics --- hence, the problem.

Of course, while seeking to unify a conflicted corpus, as a matter of practice physics typically proceeds by trying to hold onto what is believed to be any crucial, enduring insights of that corpus. In the case of QG, one obvious reconciliatory strategy begins with the observation that perhaps it is no requirement of GR that matter have a \emph{fundamentally} classical nature. Rather, perhaps, at least for the sake of phenomenological modeling, there exist an \emph{effective} classical description of the matter; or, more exactly, perhaps matter can be described by a classical stress-energy tensor. In the context of QFT it is natural to associate any such effective classical quantities with expectation values of quantum observables, where (in a Hilbert space representation of the quantum state space) the latter are modeled as operators. Thus, one arrives at the M\o ller-Rosenfeld equation, dating back to the early 1960s \citep{moller1962theories, rosenfeld1963quantization}:

\begin{equation}
\label{eq:MR1}
G_{\mu\nu}=\frac{8\pi G}{c^4}\<\hat{T}_{\mu\nu}\>.
\end{equation}
That is, the Einstein tensor $G_{\mu\nu}$, familiar from GR, couples to the expectation value of (what is now) a \emph{quantum} stress-energy tensor operator, understood to act on any given prepared state of matter. (\ref{eq:MR1}) thus modifies the Einstein field equation of classical GR, replacing the stress-energy tensor on the right-hand side with its quantum expectation value. 

It is worth stressing that, despite looking (perhaps) innocuous as a modification to the Einstein field equation from GR, the M\o ller-Rosenfeld equation is a substantive conceptual departure from the classical equation. In the first place, whereas the left-hand side features a quantity that is meant to be descriptive of a single system, the right-hand side appears to describe a statistical property of a whole ensemble of systems. To see the point, imagine a version of (\ref{eq:MR1}) in which the right-hand side is an expectation value of a classical quantity, denoted by the same brackets, but no hat: it would describe a system in which each run of an experiment had the same Einstein tensor, determined by the statistical average of the different stress-energy tensors found in each run. In other words, the geometry on any particular run would depend on the stress-energy of all past and future runs, though only those that somehow are determined to be a part of the same ensemble. The acausal structure of this statistical modification of classical GR should make it apparent that our physics is simply not like that (deep down)! But the same point would apply to (\ref{eq:MR1}) itself if we took $\<\hat{T}_{\mu\nu}\>$ as a classical expectation value over an ensemble of runs of a (quantum) experiment.

Of course in quantum theory there is a ready and standard reading of `expectation value' applicable to a single system: the sum of eigenvalues weighted by the amplitudes squared of the corresponding terms in the quantum state of that system in an individual run. While the Born Rule entails that the ensemble average will (probably) agree with that value, the quantity itself is well-defined in terms of the state of the single system, unlike the classical case. Even so, we will see in section \ref{secPageandGeilker} --- in the context of the measurement problem --- that there can be ambiguities in how we move between the classical reading of the expectation value and the quantum reading in analyses of quantum experiments.

How one constructs a quantum stress-energy tensor operator in QFT is a subtle business. But, once defined, it is indeed an operator that acts on states $|\psi\rangle$ of a material quantum system. As such, the states will obey the Schr\"odinger equation, with Hamiltonians describing both the dynamics of matter with itself, and with gravity:\footnote{\label{fnproblemoftime}Note in interpreting both  (\ref{eq:MR1}) and (\ref{eq:MRSE1}) in terms of a common notion of time, in this manuscript we sweep under the carpet the `problem of time' familiar in QG research, without further comment.}

\begin{equation}
\label{eq:MRSE1}
i\partial_t|\psi\>=\hat H_{\mathrm{matter}+\mathrm{gravity}}|\psi\>.
\end{equation} 
A system described by equations (\ref{eq:MR1}-\ref{eq:MRSE1}) is often referred to as `semi-classical gravity' (SCG), and the M\o ller-Rosenfeld equation rechristened the `semi-classical Einstein' equation. But this usage hides an important ambiguity, which can (and does) lead to significant miscommunication. On the one hand, one might take equations (\ref{eq:MR1}-\ref{eq:MRSE1}) as jointly comprising an \emph{approximation} to the dynamics of a full solution to the problem of QG, perhaps along the lines of string theory or loop quantum gravity. On the other hand, they might be proposed as a \emph{full solution} to the problem themselves: that is, where gravity is fundamentally classical, so that the quantum nature of matter entails a semi-classical theory. Let's take these two possibilities (and a third that will arise) in turn.

\subsubsection*{View 1: SCG as a mean-field description in low-energy quantum gravity}

The first reading holds that classical GR succeeds under ordinary circumstances because such circumstances reside in a regime in which it provides a good approximation to an as-yet unknown fundamental theory of QG. More particularly, models of GR are taken to be `mean-field' solutions of the unknown theory. Moreover, it is further assumed that quantum perturbations around those solutions --- `gravitons' in a broad sense\footnote{`Broad' here is opposed to the `narrow' sense of massless spin-2 quanta. The idea is that quantum perturbations can be taken around arbitrary classical solutions, while massless spin-2 particles only exist in spacetimes with suitable symmetries, for instance Minkowski and de Sitter spacetimes (see \cite[chapter 9]{huggett2020out}).} --- provide an effective field theory (EFT) for the underlying theory, something dubbed `low-energy quantum gravity' (LEQG) in \cite{Wallace}.\footnote{Note that LEQG does not comprise all that which might be reasonably termed  `low-energy quantum gravity', as for instance perturbative quantum cosmology (or even non-perturbative, yet symmetry reduced quantum cosmology in general). Note too that we leave it open whether `mean field' is to be taken more literally or more analogously, depending on the nature of the underlying QG and the limit taken to obtain the EFT.} SCG then amounts to a statement of the low-energy limit of the gravitational EFT, when fluctuations in the LEQG field may be ignored. Such a view is made compelling in part because, in a sense, merely the demand that the local spacetime metric have a Lorentzian signature in the case of quantum field theories of matter is sufficient to recover equation (\ref{eq:MR1}) to lowest order, as was historically emphasized in Sakharov's ``induced gravity'' program \citep{visser2002sakharov}. 

Indeed, given the strength of the gravitational interaction relative to others, one would expect leading order corrections to the expectation value of the stress-energy operator, for a given state of matter, to come from quantum fluctuations of the matter field itself, and not from gravitational fluctuations. Then, even though one expects deviations from classical gravity in the long-term because of the non-linear character of (\ref{eq:MR1}), for short durations of time, it is sufficient to model such a system in terms of accumulating effects of back-reaction by matter corrections on the spacetime curvature, which would otherwise determine the left-hand side of the equation. Such a modeling program is known as `stochastic gravity', and has been successfully developed (see, e.g., \citep{hu2008stochastic}). As noted by \cite[p. 22-23]{Wallace}, stochastic gravity may be derived from LEQG, indicating no tension between the present view of SCG and the expectation that stochastic noise due to the quantum nature of matter influences the effective classical description of gravity.

What is crucial to this view is that gravity is understood as fundamentally quantum in nature, and only effectively treated as classical, i.e. for the purposes of approximation. This approximation is summarized by the dictates of SCG, interpreted in terms of LEQG, perhaps improved with corrections from stochastic gravity. 

\subsubsection*{View 2: SCG as a candidate fundamental theory of gravity}

According to a second possible view, (\ref{eq:MR1}-\ref{eq:MRSE1}) constitute `fundamental' equations of motion, not approximations. This view is not taken as a serious possibility by the QG community, yet there have long been efforts to rule it out definitively: \cite{huggett2001quantize} critique theoretical arguments and also the experimental approach that we discuss below. Likely, that this view is not taken seriously in part reflects the fact that it is beset by mathematical difficulties. Namely, it is not clear that any spacetime and quantum field could simultaneously satisfy both equations. One can define a QFT satisfying (\ref{eq:MRSE1}) in a given curved background spacetime (and perhaps solve the equations), but once one has likely its stress-energy will not satisfy (\ref{eq:MR1}) in the background. One might then introduce a new background for which (\ref{eq:MR1}) is satisfied, but now (\ref{eq:MRSE1}) will likely not hold, and a new QFT must be defined. And so on. As we say, perhaps the equations can be solved simultaneously, perhaps the process described even converges on a solution (and perhaps merely approximately so), but it is far from sure. 

Still, physicists are no strangers to mathematical difficulties in the course of theoretical research. Arguably, what is more \emph{conceptually} troubling is the disunity involved in accepting that some parts of the world are classical while others are quantum even at the most fundamental level. Moreover, there are difficulties contemplating what even some basic physical models of such a theory would look like, beyond maybe the vacuum sector. For these reasons, in conjunction with the mathematical difficulties, it is perhaps not surprising that this view is given little credence by physicists. At the same time it is, for reasons to be explored later, considered worth eliminating as a live possibility.

Finally, it is not even clear that the equations (\ref{eq:MR1}-\ref{eq:MRSE1}) adequately capture what is claimed: a meshing together of classical gravity and quantum matter, as we  currently understand them. Recall in view 1 that corrections due to matter fluctuations, as studied in stochastic gravity, plausibly are relevant to the gravitational properties of a classically gravitating quantum system, so that it is a virtue of that view that LEQG recovers familiar stochastic gravity techniques. As just stated, view 2 categorically denies the role for any such corrections. The result is that there is rampant loss of physical information in coupling the material quantum system to gravity: after all, expectation values are insensitive to all higher-order correlators in the QFT. 

\subsubsection*{View 3: SCG as a mean-field description of $X$}

Now, taking SCG as an approximation does not in itself commit one to it being an approximation \emph{to LEQG}; perhaps the low energy approximation to a fundamental theory of QG is not LEQG, but some other theory, so that SCG is a ``mean field description'' of that. Now, there are few serious advocates for such alternatives, but some have been mooted: for instance, \cite{carney2019tabletop} discuss a toy model in which a mesoscopic `ancilla' continuously weakly monitors the fluctuations in the microscopic quantum fields, feeding the inevitably noisy record of those fluctuations directly into a classical gravitational interaction channel.\footnote{A version of this model was first proposed and studied by Kafri, Milburn, and Taylor in a pair of articles \citep{kafri2014classical,kafri2015bounds}, and so is sometimes referred to as the KTM model. As elaborated in \citep{altamirano2018gravity}, there are strong observational constraints on proposals for which gravity is fundamentally classical and meanwhile the pairwise gravitational interaction between macroscopic quantum sources is effectively Newtonian.} The point is not to advocate for a particular theory, but to demonstrate that one could remain \emph{neutral} on the exact relationship between QG and SCG: accept that SCG approximates some theory $X$, which in turn approximates fundamental QG (or even something wilder), but take no stance on whether $X$ is LEQG, or some theory in which gravity is classical, or some alternative, considered or unconsidered. 

\paragraph*{}

In the discussion so far, introducing view 3 may seem unmotivated; what is the point of singling out this particular epistemic stance? The vast majority of theorists seem to adopt view 1, for instance. One theme through much of the remainder of this article is that the differences across these three different views of SCG, given our current best physics, \emph{really matters} to how we understand experiments in tabletop quantum gravity. Take for instance the Page and Geilker tabletop experiment to be presented below. We will show how the extremely strong interpretation of the experimental results given by the original authors, in contrast to the typically deflationary attitude found in response to those results, reflects the difference between SCG view 1 and SCG view 3, where each is understood as opposed to SCG view 2. Getting clear on this subject will go a long way toward clarifying the terms of the disagreement regarding the significance of the proposed GIE experiments.

\section{Experimental prelude: quantum probes in two traditions}

\label{sec:experimentalprelude}

In the previous section, we explored three ways of understanding SCG, in the context of ongoing theoretical research regarding the problem of QG. Here, we discuss two experiments in the history of tabletop quantum gravity. As we will stress, the standard presentations of what is (or is not) accomplished in performing each of these respective experiments would seem to represent two different traditions in experimentation within the discipline, or two different ways of hoping to provide increasingly better empirical traction on the same problem. Following this section we will, in part, consider the new class of gravcat experiments by the lights of each of these traditions. 

\subsection{COW}

Consider first the COW experiment (named after Collela, Overhauser [\citeyear{overhauser1974experimental}] and Werner \citep{colella1975observation}): neutrons pass through a beam splitter, and the different components of their wavefunction follow paths at different heights in the Earth's gravitational field, producing a relative phase shift between the components, which can be observed as interference on recombination. Our discussion of this scenario will largely be within NRQM, but it can also be modeled covariantly to little difference, as we shall briefly see. As with all the cases we discuss, the question of whether the effect is quantum is delicate.

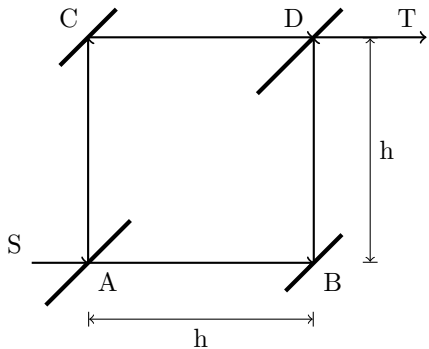
\begin{figure}[htbp]
\begin{center}

\begin{tikzpicture}[scale=.75]

\draw[ultra thick] (-.75,-.75) -- (.75,.75); 
\draw[ultra thick] (3,3) -- (4.5,4.5);
\draw[ultra thick] (-.5,3.5) -- (.5,4.5); 
\draw[ultra thick] (3.5,-.5) -- (4.5,.5); 

\draw[thick,->] (-1,0) node[anchor=south east] {S} -- (0,0) node[anchor=north west] {A};

\draw[thick,->] (0,0) -- (4,0) node[anchor=north west] {B};
\draw[thick,->] (4,0) -- (4,4) node[anchor=south east] {D};

\draw[thick,->] (0,0) -- (0,4) node[anchor=south east] {C};
\draw[thick,->] (0,4) -- (4,4);

\draw[thick,->] (4,4) -- (6,4) node[anchor=south east] {T};

\draw[|<->|] (5,0) -- (5,2) node[anchor=west] {h} -- (5,4);
\draw[|<->|] (0,-1) -- (2,-1) node[anchor=north] {h} -- (4,-1);

\end{tikzpicture}
\caption{Schematic diagram of the COW neutron interferometry experiment.}
\label{fig:COWbasic}
\end{center}
\end{figure}

The basic setup is shown in figure \ref{fig:COWbasic}, with the Earth understood to be at the bottom of the diagram: the neutron beam enters from the source S at left, is split at A, with the two components following lower and upper paths and recombining at D, with interference due to any relative phase of the components observed at T.\footnote{The geometry of the actual experiment was a parallelogram, and the device could rotate about a horizontal axis in its plane, to change the relative heights of the paths and measure the change in phase shift.} What shift should we predict? 

Consider a non-relativistic description first: Let us suppose that the neutron beam is well-described by a plane wave, $\psi(r,t)=Ce^{-i(pr-Et)/\hbar}$ as dictated by the Schr\"odinger equation, so that the phase is $pr-ET$ in $\hbar=1$ units (here $r$ is the distance along a path). Then one might suppose that the effect is the result of CD and AB being at different heights in the Earth's gravitational field, and hence corresponding to different gravitational potential energy (GPE) for each component along those path segments. But this would be a mistake. First, neutron energy is of course conserved, with changes in GPE cancelling with changes in kinetic energy (KE). (Note that in the gravcat experiments discussed later, the KE will be zero, and differences in GPE can affect the phase through the $Et$ component.) Of course, a neutron will decelerate along AC and hence take longer to traverse CD than AB, and so the lower path ABD represents less time than the upper path ACD, even though the energy is the same along both. Does this lead to a phase shift of $E\cdot(t_{ACD}-t_{ABD})$? No. Because the part of the wave reaching D along the upper path ACD takes an extra $t_{ACD}-t_{ABD}$ to get there, it must have left S a time $t_{ACD}-t_{ABD}$ earlier than the wave along lower path, in order for them to arrive simultaneously. Hence they would have been out of phase on emission by exactly $- E\cdot(t_{ACD}-t_{ABD})$, exactly canceling the difference picked up around the path.\footnote{This result is general. If two parts of a coherent wave travel along paths to a target $T$ in different times, the resulting phase shift is exactly canceled by the time difference in emission. \cite{mannheim1998classical} nicely explains how this effect is necessary for understanding the familiar two-slit experiment.} In other words, the $Et$ part of the phase contributes no relative phase, and the whole observed effect is due to the $pr$ part of the phase. Before we go on to calculate this quantity, note that when considering the deceleration of neutrons due to gravity, we have treated them as classical particles. Indeed, the entire calculation of $kr$ will proceed in this way, with implications for the quantum nature of the phenomenon. 

The original OW \cite{overhauser1974experimental} prediction is based then on the lower momentum along CD than AB, because of gravitational deceleration along AC. Basic mechanics tells us that a particle of mass $m$ and initial vertical speed $u$ will have a final speed $\sqrt{u^2-2gh}$ after traveling a distance $h$; or to first order in $g$, a speed of $u-gh/u$. Assuming that the reflection is perfect, the phase change along $CD$ is thus $m(u-gh/u)h$, while that along $AC$ is simply $muh$ since no deceleration is involved. The momenta along the vertical legs should be the same, and so OW predict a relative phase shift of $mgh^2/u$, which is indeed observed.\footnote{In fact the observed value was around 10\% lower. The largest correction comes from  deformation of the crystalline structure of the apparatus, which is carved from a single crystal of silicon. Once this is accounted for, other effects can be measured by further deviations from the prediction, including that due to the earth's rotation. See \cite{greenberger1980role}.}

However, since OW use classical particle considerations to calculate the motion of the neutron beam, they should do so consistently, for instance taking into account that paths will not be horizontal but parabolic as the neutrons fall during their flight: they reach the plates not at B and D, but slightly lower (see figure \ref{fig:COWMann1}). This perturbation affects the time, magnitude, and direction of the paths, and at first order so must be accounted for. OW reference this effect, but the analysis of Mannheim shows that they do not correctly compute it: indeed, he shows that there is \emph{no} relative phase along the closed part of the paths! Instead, the particle considerations that OW invoke, when applied consistently, reveal a different source for the observed interference --- which happens to agree with their original computation! (With hindsight, the coincidence is perhaps not a great surprise given that $mgh^2/u$ is the natural dimensionless quantity in the problem.)\footnote{\cite{greenberger1979coherence} show that under certain assumptions the effect of parabolic motion is (to lowest order) \emph{numerically} equivalent to integrating the potential difference along the paths, and thereby also derive the same result. However, they do not take into account all the factors that Mannheim does: especially the change in momentum and reflection angle caused by parabolic motion. So they do not establish the full equivalence of their method and Mannheim's analysis.}

The details of Mannheim's calculation are somewhat off the main line of this manuscript, though illustrative of how one might theoretically reconcile classical gravity and quantum matter in such a situation: hence they are found in an appendix.\footnote{Those worried here about calling gravity `classical' when it must appear as an operator to act on quantum matter may be reassured in this particular case by the following: in the covariant treatment of the experiment discussed below, there is an equivalent prediction in strictly \emph{classical} optics.} The bottom line is, first, that there is no phase difference at all between a point on the splitter and the recombining screen along the two paths. However, second, a point on the wavefront at the splitter will not reach the same point on the recombining screen (see again figure \ref{fig:COWMann1}); rather, wavefront points that are slightly offset on the splitter reach the same point of recombination. Finally, since the splitter is not perpendicular to the neutron beam, such wavefront points will travel different distances to the splitter (see figure \ref{fig:COWMann2}), and so be out of phase, \emph{completely accounting for the effect}!

One might thus conclude that the effect is not gravitational at all! But this would be too hasty: although the phase changes along the paths cancel, their explanations involve two quite different combinations of speed and trajectory shifts, and the Earth's gravity is the cause in both. Indeed, in the actual experiment, when the apparatus is rotated about a horizontal axis through its plane, so that the `vertical' paths are no longer vertical, and the horizontal paths change their relative heights, the phase changes no longer cancel, and the observed relative phase changes.

As an aside to our main topic, we note that observing a phase shift $mgh^2/u$ amounts to a measurement of the neutron mass, since the size of the experiment and the speed of the neutron are known. This is remarkable, because in the vicinity of the equivalence principle (EP) is the idea that it is impossible to measure the mass of a body using purely gravitational effects --- but that is exactly what is accomplished by the COW experiment.\footnote{This point was made in \cite{okon2011does}.} Indeed, one might therefore worry that the experiment reveals an incompatibility between the EP and NRQM. However, reflection on the equation of motion, the Schr\"odinger equation, of course reveals the identity of gravitational and inertial mass that constitutes the Newtonian EP: the same mass $m$ appears in the kinetic term that governs inertial motion, and in the potential term that determines the gravitational force. (The same points hold, mutatis mutandis, in a covariant analysis \cite[p. 57]{mannheim1998classical}.)\footnote{\cite{brown1996bovine} considers other implications of the experiment, especially for the action-reaction principle.} 
What though does the experiment show about the relation between Newtonian gravity and NRQM? Clearly the effect is quantum in the sense that in the $\hbar\to0$ limit the neutron is purely classical, and has no phase at all. Additionally, in Mannheim's analysis, the effect depends on the finite spatial (and temporal) spread of the wavefunction. However, our question regards the nature of the \emph{interaction} of gravity with quantum matter, the neutrons in this case. So a more relevant consideration is that the GPE has no direct effect on the phase, as one might have naively thought: as we discussed, the energy is conserved (and the time from source to target is irrelevant) so we find no $Et$ component of the phase. Moreover, the calculations of $k$ and $r$, which do contribute to the phase, depend on purely classical properties of the neutrons: their trajectories determined by Newtonian mechanics in a constant gravitational field. So in that regard, gravity and matter are related purely classically in the experiment. (Note that all these considerations -- except the finite spatial extent of the wavefunction -- apply equally to OW's analysis.)

As we mentioned earlier, we have focused on a NRQM model of the experiment, in order to simplify a fairly complex situation. However, Mannheim also outlines a covariant model, which it is worth briefly discussing to see that it produces the same result, and also as a warm-up for later calculations. First, for the metric describing the Earth's gravitational field we use the weak field approximation to the Schwarzschild solution to the Einstein field equation:

\begin{equation}
\label{eq:WF}
ds^2=(1+2V(\vec r)/c^2)dt^2 - d{\vec r}^2,
\end{equation}
where $V(\vec r)$ is the Newtonian potential, in this case $-gx$ when the Earth is the source of the field. Timelike geodesics then satisfy $\ddot y=\ddot z=0$, $\ddot x=g$, so that in this approximation neutrons will follow the same paths that we computed in the Newtonian analysis. Moreover, treating the neutrons as excitations of a Klein-Gordon quantum field reveals that the phase change is again dependent on the momentum along each path. Hence over all, Mannheim's Newtonian calculation of neutron interference applies equally to the covariant treatment.

That said, the covariant situation is different, insofar as the effect also has an entirely classical realization, unlike a Newtonian analysis. This is because covariantly light will follow null geodesics, and not the straight and vertical paths expected in a Newtonian treatment. Remarkably, the very same considerations that apply to massive particles apply to massless classical waves: interference is predicted for a classical light beam COW experiment, for exactly the same reasons as for the quantum neutron experiment. Thus one can envision a gravitational Michelson-Morley experiment to observe the bending of light in the Earth's gravitational field!\footnote{We thank Professor Mannheim for discussions of these matters. He reports that an apparatus with arms 1km long would be capable of measuring the effect: in stronger fields of course such an apparatus would only need shorter arms to measure the bending of light.} Thus there is a purely classical realization of the effect covariantly, though not in the Newtonian treatment.

Since one expects the analogue of the COW experiment even in this classical system, it seems that the COW experiment's credentials as an observation of the quantum nature of gravity are at best weak. But looking at things in this way misses the main point of the experiment. Instead, one is interested in treating the gravitational interaction channel as opaque: to be studied with increasingly sophisticated quantum probes, in terms of the effects registered in those probes. In doing so, one hopes that physicists might develop good experimental \emph{control} over the gravitational behaviors of material quantum systems that are increasingly poorly described as classical, providing new knowledge (a `causal manipulation' kind of knowledge) that is relevant in the course of theoretical research on the problem of QG. So, for instance, one may study the dynamics of individual quantum probes isolated from their (known) quantum environments, but which are nonetheless subject to gravity for the duration.

\subsection{Page and Geilker}\label{secPageandGeilker}

The second experiment is that conducted by \cite{page1981indirect}, and reported  as `indirect evidence for quantum gravity'. So begins, in contrast with what we have just said about COW, the tradition of experiments in tabletop quantum gravity that claim to search for increasingly sophisticated \emph{witnesses} of the underlying quantum nature of gravity.

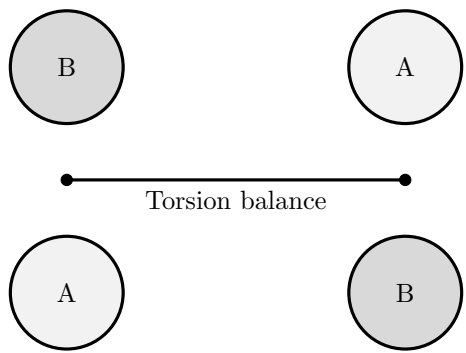
\begin{figure}[htbp]
\begin{center}

\begin{tikzpicture}[scale=.75]

\filldraw[fill=gray!30, very thick](-3,+2) circle (1) node[] {B} ;
\filldraw[fill=gray!10, very thick](-3,-2) circle (1) node[] {A} ;
\filldraw[fill=gray!10, very thick](3,+2) circle (1) node[] {A} ;
\filldraw[fill=gray!30, very thick](3,-2) circle (1)  node[] {B} ;

\draw[very thick](-3,0)--(0,0)node[anchor=north] {Torsion balance}--(+3,0)   ;
\filldraw[](-3,0) circle (0.1);
\filldraw[](3,0) circle (0.1);

\end{tikzpicture}
\caption{Schematic diagram of the quantum Cavendish experiment from above. A and B represent the two possible locations of the weights; on each run the actual position is decided by the reading of a Geiger counter.}
\label{fig:P-G}
\end{center}
\end{figure}

The Page and Geilker experiment is a modification of the famous Cavendish experiment. In the original, two smaller test masses are placed on the ends of a hanging torsion balance, and a larger source mass is placed close to each, but on opposite sides of each end: the gravitational attraction between the test and source masses is thereby measured by the deflection of the balance. If the weights are placed in, say, position A in figure \ref{fig:P-G}, then the balance will deflect counterclockwise. In the authors’ modification, quantum mechanical radioactive decay is used to randomize between two classical configurations: the weights at A exerting a counterclockwise torque, and the weights at B a clockwise torque. Since radioactive decay is a quantum process, \citeauthor{page1981indirect} reason that in each run the weights will be in a superposition of classical positions A and B. 

Of course, the \emph{expected} position of the weights in such a state is between the two positions, co-located with the test masses, and so SCG --- which depends on the expected distribution of matter --- predicts that no net gravitational force is exerted on those test masses, and the torsion should vanish in every run. However, (unsurprisingly) in their experiment \citeauthor{page1981indirect} observed that the balance always deflected, with equal frequency in each direction: although SCG predicts the correct expected displacement, it is wildly incorrect with respect to the individual runs. (The observed result is of course compatible with an entirely different set-up, one in which the source weights are \emph{randomly} prepared in each trial in just one of the two classical configurations. So if the experiment is to rule out SCG, it is crucial that the choice of position is determined quantum mechanically.)

In the first place, Page and Geilker intend that the experiment serve as a refutation of SCG, construed as a candidate fundamental theory of gravity, as in view 2. As such, the result is of course unsurprising --- we do not expect macroscopic source mass positions to coherently superpose!\footnote{One might perhaps try to read the original conclusion of \cite{page1981indirect} as a refutation of SCG, understood on either SCG view 1 or SCG view 3, as providing an apt approximation scheme in the specific physical domain of the experiment. This is right, of course (as we will discuss presently), but it seems to us a stretch to imagine that this is what the authors had in mind.} (Thinking of our earlier discussion of the M\o ller-Rosenfeld equation specifically, we expect the torsion in any one run to depend on the position of the sources in that run, not their average position over a series of runs.) However, caution is merited. As the experiment depends on the outcomes of individual runs rather than a statistical average for an ensemble, one simply cannot interpret the experiment as a test of SCG without some answer to the `measurement problem' in mind.\footnote{\cite{EmilyAdlam} offers a nice discussion on how tabletop quantum gravity experiments relate to questions on interpretations of quantum mechanics, including the present experiment as well as the new GIE experiments discussed centrally below (which are, in fact, her focus as well).}

Indeed, Page and Geilker assume an Everettian interpretation of quantum mechanics (QM): the experiment is a massively open system --- the sources are placed in position by a (human) experimenter who observes the outcome of the decay. Thus the A-placement and B-placement branches decohere, and the observed deflections come from a random sampling of the branches over an ensemble of experiments. Granted then, the experiment refutes SCG as a candidate fundamental theory, given an Everett interpretation.

Of course, standard interpretations of quantum mechanics predict the same outcome if SCG fails. As \cite[p.11-12]{Wallace} points out, if the interpretations use only unitary dynamics, then they will utilize the same decoherence argument as Everettians. While if an interpretation invokes some kind of collapse, real or ersatz, it will occur when the experimenter measures whether a decay occurred as part of the procedure, `collapsing' the system into one of the branches. Thus, no one would plausibly expect a macroscopic superposition of the source masses in the experiment, so at the macroscopic scales relevant here, classical descriptions of (fundamentally) quantum physics suffice to capture, \emph{in each run}, the gravitational dynamics of the experiment.

However, for this very reason, interpretations that go beyond unitary dynamics may also predict the same outcomes even if SCG is true. For example, since collapse interpretations such as that of Ghirardi, Rimini and Weber (\citeyear{ghirardi1986unified}) entail a collapse once the decay (or its absence) is observed by the experimenter, they predict that the source masses are never in superposition: so the expected mass distribution is always either in position A or in position B. Similarly, a hidden variable theory, such as Bohm's, could trace a definite configuration to some uncontrollable --- hence effectively random --- epistemic uncertainty in the initial state, leading to `effective collapse' \cite[\S3.2]{durr1992quantum}. In other words, the implications of Page and Geilker's experiment are sensitive to the interpretation of QM: according to the Everett interpretation it does serve to rule out SCG, while for collapse or hidden variables it may not do so. One would like to close such loopholes; the gravcat experiments to be discussed later would achieve this goal (and maybe rather more).\footnote{However, we concur with \cite{Wallace} that physicists who claim that standard quantum dynamics is complete really \emph{should} accept the Page and Geilker experiment as experimentally refuting view 2.}

Page and Geilker not only take --- according to their Everettian lights --- the experiment to refute view 2 of SCG, they also take it to provide indirect evidence for the quantum nature of gravity. Rightly so: our posterior credence conditional on the evidence gathered gets spread across the remaining alternatives, including SCG view 1: SCG as a mean-field description in LEQG. Now as the present discussion showcases, whether this inference is apt depends at least in part on attitudes towards quantum measurement. But meanwhile, how seriously we take the refutation of view 2 to count as evidence for the quantum nature of gravity also depends on downplaying the possibility that mesoscopic processes can screen out microscopic effects of quantum matter in a macroscopic gravitational experiment, namely what is allowed on view 3. On this understanding of SCG, the observed classicality of the gravitational interaction within the individual runs of their experiment would not necessarily mean that gravity is fundamentally quantum. On the other hand, absent cause to think that nature conspires to keep gravity fundamentally not quantum, such a refutation of SCG does seem to point towards the authors' ultimate conclusion. But by the same token it is clear why many physicists' reactions to the experiment were not enthusiastic (even assuming the Everett interpretation). To the extent that the upshot of the experiment is to have in some sense produced indirect observational support for quantum gravity --- to have  `witnessed' the underlying quantum nature of the gravitational state --- the added value of performing the experiment depends rather sensitively on the prior plausibility one assigns to SCG view 1 over view 3. Namely, since the prior probability for view 2 is, in light of the experiment, redistributed over the other two views in proportion to their priors, unless one already takes view 1 to be considerably more probable than view 3, then the probability for quantized gravity could not increase significantly.\footnote{Note that, as defined, the support in theory space of views 1 and 3 respectively are not mutually disjoint, as view 3 was neutral where view 1 took a stand. To the extent that one might imagine the distribution of credences to be defined over theory space (i.e. rather than the set of the three views equipped with the counting measure), the above should not be read as asserting that the prior probability for view 2 is redistributed across views 1 and 3 relative to their priors, in the manner of a convex sum. But the general point goes through in either imagined case.}

Thus, whether one already has a low prior for view 2, or has comparable priors for views 1 and view 3, or adopts a collapse interpretation of QM, one is indeed likely to agree with \citeauthor{ballentine1982comment}: indeed, ``a less surprising experimental result has scarcely, if ever, been published.''

\section{Gravitationally Induced Entanglement experiments}
\label{secGIE}

The COW experiment, while showing that gravitational fields affect quantum matter, considers only the field of a classical source, the Earth. The Page and Geilker experiment claims to involve the gravitational fields of sources in quantum superposition (though decohered), but its interpretation as evidence of quantum gravity involves loaded assumptions. No wonder then, that many have looked for more clear cut demonstrations of the `quantum nature of gravity'. And so we turn to the main object of this manuscript: recently proposed experiments that arguably can demonstrate the quantum nature of gravity, and thereby present as a new frontier in the witnessing tradition that began with Page and Geilker.\footnote{\label{fnCMB}\cite{Wallace} points out that the explanation of cosmic microwave background (CMB) structure in terms of fluctuations in the inflaton field in the early universe requires gravity to be a quantum field (\S6). One might consider this data to be a prior `witness' of the quantum nature of gravity on this view. We note that this conclusion requires considerably more theoretical assumptions than the GIE experiment (leaving open a strong sense in which the GIE experiment aspires to be a prior witness after all): not only the standard theory of inflation, but also that classical gravity does not couple to higher order matter fluctuations. (That is, that stochastic gravity cannot equally explain the CMB data, as for instance argued in \cite{RouraVerdaguer}.) So the GIE experiments would be an advance both by exploring a new regime, and by its relative theoretical neutrality. (Though as we noted, we will see it is not completely neutral by any means!)}

\subsection{Direct or indirect?}

In the following subsection, we will give a first account of these new `tabletop' experiments. They centrally feature a `gravitational Schr\"odinger cat', or `gravcat' (as a neologism has it) --- an uncharged object large enough to exert a gravitational force greater than any van der Waals forces, but small enough that it can be placed and preserved in a quantum superposition for long enough to interact with another gravcat and then have their entanglement measured. It has recently been proposed by Bose et al (\citeyear{bose2017spin}), and Marletto and Vedral (\citeyear{MarlettoVedral}) that a pair of gravcats could be used to probe the quantum nature of gravity, in experiments that could in principle, and perhaps in practice within the next decade or so, be carried out on a lab bench (perhaps using $10^{-14}$kg diamonds). Describing what has become known as a `GIE experiment',\footnote{Names may not be entirely settled; GIE here stands for ``Gravitationally Induced Entanglement''. To emphasize the novelty, we may regard the COW experiment as a Gravitationally Induced \emph{Interference} experiment.} Marletto and Vedral write: ``the entanglement between the positional degrees of freedom of the masses is an indirect witness of the quantization of the gravitational field." That is, the observation of the quantum nature of gravity would be rather \emph{indirect}. Crucially then, such experiments are distinct from --- but far easier than --- \emph{direct} observations of the quantum nature of gravity, e.g. from the observation of individual gravitons or their effects, for instance, some gravitational analogue of the photoelectric effect .\footnote{\label{ftnt:GGE}Observing a strict analogue to photoelectric effect is fanciful: for the $n=2$ to $n=1$ hydrogen transition the electrical potential energy is $~10$eV, but the gravitational potential energy is $~10^{-38}$eV, so while the former is readily observable, the later is not (to say the least!). More practical --- but still impossible by many orders of magnitude --- is the observation of Gravitationally Induced Decoherence of entangled systems (see \cite[\S 3.2]{carney2019tabletop}), though arguably this too would be `indirect' evidence.}

Let us clarify what we mean by the `direct/indirect' distinction, as the literature appears to show some cross-talk in this regard. Namely, \citeauthor{MarlettoVedral}'s classification of the experiment as \emph{in}direct (also \cite{carney2019tabletop}) is at odds with the discussion found in the supplementary materials of \cite{bose2017spin}, where one reads: ``it is fair to say that there are no feasible ideas yet to test [QG]'s quantum coherent behavior directly in a laboratory experiment. Here, we introduce an idea for such a test...''. Likely, this clash is terminological. For instance, \cite[p. 65]{christodoulou2019possibility} cite both as supporting the same conclusion: ``As emphasised by Bose et al. and by Marletto and Vedral, the main reason for the interest of the experiment is precisely to provide \emph{direct} evidence that gravity is quantised" (our emphasis).\footnote{Exactly what is the observed quantum gravitational system is not the same across their theoretical treatments: for Marletto and Vedral gravity is an information channel, for Bose et al a quantum field theory state, and for Christodoulou and and Rovelli a superposition of metric geometries. Of course, ultimately they all take these  to be different representations of the same object.}

So perhaps there is consistency across these claims, but there still seem to be some competing intuitions about how to draw a direct/indirect distinction, and how `witnessing' might relate to them. The philosophical inclination is of course to attempt a precisification, but we have found that so doing does little to clarify the questions and debates we engage, so we will simply put aside all such questions by \emph{stipulating} that these authors intend the experiment as an `indirect witness' of the quantum nature of gravity. Doing so does not diminish the task at hand --- namely, of evaluating the stakes of the new experiment from within the witness tradition. Even adopting this definition, we will soon see that there remains a substantive question of whether the experiment could witness any such thing as the quantum nature of gravity at all. (Indeed, in following subsections we will discuss different models one might have of the experiment on the basis of consideration of all the relevant fundamental physics, and how those different models might lead to different answers to this question, showing that it is not an unequivocal matter.) 

We will return to the question of what to make of GIE experiments beyond witnessing in section \ref{secMakingQuantum}. For now, with the previous discussion in mind, let us turn to a first, naive pass at the experiment itself, to get the basic ideas in hand.

\subsection{A naive account of the experiments}

In the experiment, two identical gravcats are prepared in position superpositions along the x-axis. Let us say that the first gravcat is the sum of gaussians located at $-D\pm\Delta$, while the second is a superposition of gaussians located at $+D\pm\Delta$; let us set $D\gtrsim\Delta$. Given their macroscopic size, we can take the gravcats to have no initial motion along the x-axis. Moreover, provided that the duration of the experiment is short, the gravcats will not have time to accelerate due to mutual gravitational attraction; they are effectively stationary. The arrangement is shown in figure \ref{fig:gravcat}, and is described by the following state (ignoring overall normalization):
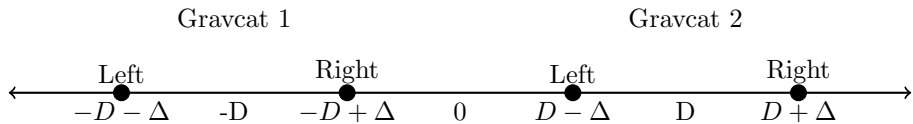
\begin{figure}[htbp]
\begin{center}
\begin{tikzpicture}[scale=1.5]

\draw[thick,<->] (-4,0) -- (0,0) node[anchor=north] {0} -- (4,0);

\filldraw[black] (-3,0) circle (2pt) node[anchor=south] {Left} node[anchor=north] {$-D-\Delta$};
\filldraw[black] (-1,0) circle (2pt) node[anchor=south] {Right} node[anchor=north] {$-D+\Delta$};
\filldraw[black] (1,0) circle (2pt) node[anchor=south] {Left} node[anchor=north] {$D-\Delta$};
\filldraw[black] (3,0) circle (2pt) node[anchor=south] {Right} node[anchor=north] {$D+\Delta$};

\filldraw[black] (-2,0) circle (0pt) node[anchor=north] {-D};
\filldraw[black] (2,0) circle (0pt) node[anchor=north] {D};

\draw[black] (-2,.5) circle (0pt) node[anchor=south] {Gravcat 1};
\draw[black] (2,.5) circle (0pt) node[anchor=south] {Gravcat 2};

\end{tikzpicture}
\caption{The two gravcat arrangement (the central packets are closer than shown).}
\label{fig:gravcat}
\end{center}
\end{figure}
\begin{equation}
\label{eqn:BS}
\Psi(0) = (|L\>+|R\>)\x(|L>+|R\>) = |L\>\x|L\> + |L\>\x|R\> + |R\>\x|L\> + |R\>\x|R\>,
\end{equation}
with $|L\>$ representing the left wavepacket of gravcat 1 in the first tensor product slot, and the left wavepacket of gravcat 2 in the second slot, and so on. 

Because the distances between the packets of the two gravcats are different, different terms in (\ref{eqn:BS}) will have different gravitational potential energy (GPE), and so will develop relative phases.\footnote{In this initial presentation of the experiment, we will continue to treat gravity as Newtonian, but we emphasize that \cite{bose2017spin}, who we follow here, argue that the effect should be understood in terms of coherent states of a quantized relativistic field. We will discuss this point in \S\ref{sec:Bose} below.} For example, for $|L\>\x|L\>$ the GPE will be $Gm^2/2D$, whilst for $|L\>\x|R\>$ it will be $Gm^2/(2D+2\Delta)$. Since the GPE is the only contribution to the energy (kinetic energy is zero), these different potentials will produce a different phase in each term according to the Schr\"odinger equation: $\Phi(t)=e^{-iEt}\Phi(0)$ for energy eigenstates such as these (recalling that we set $\hbar=1$). By linearity then:

\begin{eqnarray}
\label{exact}
\nonumber\Psi(t) = e^{\frac{-iGm^2t}{2D}}|L\>\x|L\>  +  e^{\frac{-iGm^2t}{2D+2\Delta}}|L\>\x|R\>  &+&\\
e^{\frac{-iGm^2t}{2D-2\Delta}}|R\>\x|L\> &+& e^{\frac{-iGm^2t}{2D}}|R\>\x|R\>.
\end{eqnarray}

\noindent Or, since $D\approx\Delta$, for short times ($t\ll 2D/Gm^2$)

\begin{equation}
\label{eq:Bapprox}
    \Psi(t) \approx |L\>\x|L\>  +  |L\>\x|R\>  +
e^{\frac{-iGm^2t}{\delta}}|R\>\x|L\> + |R\>\x|R\>,
\end{equation}
with $\delta=2(D-\Delta)$ the separation of the closest pair in (\ref{eq:BSspin}). This wavefunction does not factorize except for special values of $t$ (and neither does the un-approximated wavefunction \eqref{exact}), and so the mutual gravitational attraction induces entanglement between the gravcats. It is this entanglement that is claimed to provide an indirect witness of the quantum nature of gravity.\footnote{\label{fn:spin}In a little more detail, \cite{bose2017spin} propose that the gravcats have spin. Prior to the experiment they are in a state
\begin{equation}
\label{eq:BSpre}
|X_1\>(|\up\>+|\dn\>)\x|X_2\>(|\up\>+|\dn\>),
\end{equation}
where $X_1$ and $X_2$ are the initial positions of the two gravcats, and $|\up\>$ and $|\dn\>$ positive and negative x-spin states. To prepare the state (\ref{eqn:BS}) the gravcats are passed through a Stern-Gerlach device oriented along the x-axis, to produce
\begin{equation}
\label{eq:BSspin}
(|L\up\>+|R\dn\>)\x(|L\up>+|R\dn\>)=|L\up\>\x|L\up\> + |L\up\>\x|R\dn\> + |R\dn\>\x|L\up\> + |R\dn\>\x|R\dn\>.
\end{equation}
After gravitational entanglement is produced, the gravcats are passed back through the Stern-Gerlach device to yield
\begin{equation}
|X_1\up\>\x|X_2\up\> + |X_1\up\>\x|X_2\dn\> + e^{\frac{-iGm^2t}{\delta}}|X_1\dn\>\x|X_2\up\> + |X_1\dn\>\x|X_2\dn\>,
\end{equation}
using the same approximation as (\ref{eq:Bapprox}). This state exhibits spin entanglement between the gravcats, which could be observed by measurements of spin-correlations between the particles that violate Bell-type inequalities. Note that the gravcat interaction itself does not involve their spins, only their mutual gravitational attraction.} All of this is still beyond current technology, but not so far beyond that experimentalists aren't attempting to develop existing techniques to make such a measurement, or one along similar lines. 

Now, given our description of the experiment, the claim that observing such entanglement witnesses the quantum nature of gravity should seem puzzling: after all, it appears that we simply appealed to classical gravity, just as in the COW experiment (whether by their lights or Mannheim's). Yet, \cite{colella1975observation} did not claim that the neutron interference which was observed provided a witness of the `quantum nature' of gravity\footnote{Though \cite[p. 76]{greenberger1980role} makes a very weak claim in this direction.} --- indeed, we even suggested that the experiment as self-described ultimately represents work in a different tradition of experiment, one that instead emphasizes \emph{control}. How is the present appeal to classical gravity relevantly different? Of course in the COW experiment the source of the field was considered as classical, namely the Earth; while in the GIE experiment it is quantum, the gravcats themselves. But that seems a point about the gravcats, not gravity. (The same of course can be said of the sources in the Page and Geilker experiment, noting that gravcats remain in coherent superposition.)

Unpacking this issue is a major task of the remainder of the manuscript. We will see how different theoretical starting points can lead one to different conclusions. For instance, the naive analysis tacitly assumes a bipartite state of two gravcats, acting directly on each other, through an interaction Hamiltonian
\begin{equation}
\label{eq:BoseH}
\hat H = \frac{Gm^2}{|\hat x_1-\hat x_2|},
\end{equation}
in which gravity appears as a classical potential (except, of course, it is, formally, an operator).\footnote{Of course in our scenario the kinetic term is ignored, because the particles are stationary. Indeed, no contributions to the Hamiltonian that depend only on the state of just one particle --- for instance a potential in an external field --- will affect entanglement, since they act only on one part of the tensor product. Only a term that depends on the state of both particles --- a mutual potential --- will produce entanglement.} But the claim that the experiment would witness the quantum nature of gravity requires that it is something about the quantum nature of whatever the classical potential in (\ref{eq:BoseH}) actually describes that ultimately explains the predicted outcome. What we will now work out explicitly is that whether or not one takes this view --- namely, whether the ultimate explanation for the predicted outcome of the experiment concerns facts about an underlying gravitational field, or (just) the classical potential, fashioned into an operator --- amounts to a metatheoretical choice of modeling framework, or `paradigm' in a lite sense: no rational argument alone can make the proponent of one paradigm swap to the other.

To foreshadow: we will present two paradigms relative to which one can understand the gravcat experiments. The first paradigm starts from principles of our best current physics to vindicate the naive view: that is, it turns out not to be so naive after all. We shall see that there is at least one sense in which a non-local interaction may be seen to fall directly out of the fundamental field dynamics of GR, within the context of these gravcat experiments. In this paradigm then, gravcats do not seem to witness the quantum nature of gravity at all. The claim to witness comes within a second paradigm, spurred by nearly 200 years of physics, including relativity, telling us that bodies do not act at a distance on one another, but rather their interactions are mediated by fields (cf. \cite{hesse2005forces}). Exactly how to understand the gravitational field is of course equivocal, and indeed here it will mean different things: in some contexts it means the metric field of GR, in others the spin-2 massless field that appears in the linearization of GR on a Minkowski background. But at the simplest level, to treat gravity as itself a dynamical system responsible for mediating between bodies is to model the gravcat experiments as tripartite, rather than bipartite (as in the naive view), with the field joining the two gravcats in the labeling of the \textit{full} quantum state.

However, before discussing further how claims that gravcats witness the quantum nature of gravity are dependent on the paradigm adhered to, we will briefly rehearse an uncontroversial yet significant consequence of the gravcat experiment: namely, that it would rule out semi-classical gravity, as understood on view 2 in \S\ref{sec:prelude}, and in a manner less dependent on the interpretation of QM than could be claimed by \citeauthor{page1981indirect}. (Though this is not to say that the experiment makes paying attention to interpretations of QM obsolete in tabletop quantum gravity. As \cite{EmilyAdlam} argues at length, both achieving the hypothesized results and persistently failing to achieve them might be taken as relevant to the project of weighing the comparative merits of such interpretations on ultimately empirical grounds.)

\subsection{Ruling out semi-classical gravity}\label{sec:rulingoutsemiclassical} 

Quite simply, the observation of post-experiment gravcat entanglement would be in direct experimental conflict with semi-classical gravity, defined by the equations (\ref{eq:MR1}-\ref{eq:MRSE1}), taken as a fundamental theory of nature, because such an interaction cannot result in entanglement.\footnote{Putting aside the controversial Page and Geilker experiment, we know of only one other observation claimed to show the same thing, namely the explanation of Cosmic Microwave Background structure in terms of fluctuations in the inflaton field, in that model of the early universe (\citep[\S6]{Wallace}). We will return to this point in \S\ref{secMakingQuantum}.} Let us explain.

To simplify matters, let us work in the Newtonian limit of equation \ref{eq:MR1}, in which the stress-energy of each gravcat is merely its mass density, $m|\psi(x)|^2$; for each gravcat a mass density zero almost everywhere, but with equal small gaussian peaks at the locations of the two wavepackets. The effect of such a distribution is \cite[\S1.2]{anastopoulos2014problems}\footnote{They take equation (\ref{eq:MR1}) as effective; here we are taking it as fundamental, and thus their conceptual critique of the derivation does not apply. Note that \cite[\S 1.3]{HuBook} show that the Newton-Schrödinger equation does \textit{not} arise in the combined weak field limit of GR and the non-relativistic limit of quantum field theory. Instead, what arises is the equation discussed in the next subsection under the label of `Newtonian Model'.} to introduce a potential into the Hamiltonian of the form:

\begin{equation}
\label{eq:HNS}
\hat H_{\mathrm{gravity}}=-Gm^2\int dx'\frac{|\psi(x)|^2}{|x-x'|}.
\end{equation}
Inserting this term into equation (\ref{eq:MRSE1}) yields what is known as the `Newton-Schr\"odinger equation'. This approach has often been criticized, and while not plausible, arguably has not been refuted empirically; hence it remains to date as a possible theory of quantum matter and classical gravity. However, we can readily see that witnessing entanglement in the proposed way would amount to direct empirical evidence against the theory.

Since gravcats only interact gravitationally, $\hat H_{\mathrm{matter}}$ is merely the kinetic energy, which vanishes for stationary particles. Hence, assuming that the experiment does not last long enough for an appreciable change in velocity, equation (\ref{eq:MRSE1}) will only introduce phases according to this gravitational potential, which we can see will not produce entanglement. Suppose that the potential, implicit in (\ref{eq:HNS}), in which the left packet of the first gravcat sits is $V_{L1}$, and then for an initial state $|L\>$, the time-dependent Schr\"odinger equation dictates that the state is given by:

\begin{equation}
\psi_{L1}=e^{-iV_{L1}t}|L\>,
\end{equation}
and similarly for the other packets. Then, by linearity, if the initial state is again (\ref{eqn:BS}), then the Schr\"odinger equation yields as the time-dependent state:

\begin{eqnarray}
\nonumber \Psi(t) = e^{-iV_{L1}t}|L\>\x e^{-iV_{L2}t}|L\> & + & e^{-iV_{L1}t}|L\>\x e^{-iV_{R2}t}|R\>\\
 & + & e^{-iV_{R1}t}|R\>\x e^{-iV_{L2}t}|L\> + e^{-iV_{R1}t}|R\>\x e^{-iV_{R2}t}|R\>\\
\nonumber & = & (e^{-iV_{L1}t}|L\> + e^{-iV_{R1}t}|R\>)\x(e^{-iV_{L2}t}|L\> + e^{-iV_{R2}t}|R\>),
\end{eqnarray}
so that the wavefunction factorizes, and there is no entanglement. The explanation is obvious: if there is no interaction potential, then the gravcats evolve independently, and their states will preserve factorization. Hence the tabletop experiments clearly offer an empirical test of the quantum nature of gravity in so far as they rule out a classical treatment in the form of SCG, if gravcat entanglement is observed. (Though of course SCG may remain a good effective theory in suitable situations).

That said, some puzzle remains about the enthusiasm of the community for carrying out the experiment. After all, it will require considerable effort and expense, and it is almost universally believed in the community that view 2 of SCG is in fact false; why go to all that trouble to confirm something to which you already ascribe a high degree of credence? We will return to this question in section \ref{secMakingQuantum}.

\section{Two paradigms from fundamental physics}\label{sec:paradigms}

In this subsection, we present the two competing paradigms (in the lite sense) which are, on our view, decisive for whether or not one takes the gravcat experiments to witness the quantum nature of gravity: we will first present the paradigm upon which the experimental set-up has to be modeled as bipartite --- there is no witnessing of a quantum nature of gravity on this view; in the following subsection, we will then discuss the `tripartite' paradigm, which does seem to support the claim that gravcat experiments witness the quantum nature of gravity.

\subsection{The Newtonian model}

\label{sec:Newtonianmodeling}

We have already seen that, on a naive view, the Hamiltonian (\ref{eq:BoseH}) will produce entanglement, without assigning quantum states to the gravitational field. As we noted, such a view treats gravity as a direct, instantaneous interaction-at-a-distance; if one takes seriously finite relativistic propagation, and indeed the dynamical nature of the gravitational field in GR, then such an interaction cannot be allowed strictly speaking. But one might come at things from a particular, rather different starting point, and conclude that the naive view is vindicated. Here, the thought is that the treatment of Newtonian gravity as a direct, instantaneous action-at-a-distance interaction, as on the naive view, is merely an apt \emph{stand-in} for the `true' (more on this below) physical degrees of freedom, given that the experiment is assumed to take place entirely in the Newtonian regime. Specifically, \cite[\S3.1]{anastopoulos2014problems} starts with the GR action for a scalar field

\begin{equation}
S[g,\phi]= \frac{1}{8\pi G}\int dx^4 \sqrt{-g}\Big(R - \frac{1}{2}(\nabla\phi)^2 - \frac{1}{2}m^2\phi^2\Big),
\end{equation}
where $R$ is the Ricci scalar, and $\nabla$ the covariant derivative. They take a 3+1 Minkowski background, assume linear perturbations, and work in the Hamiltonian framework --- all the while imposing standard gauge constraints --- to obtain the Hamiltonian

\begin{eqnarray}
\nonumber H=\frac{1}{2}\int d\mathbf r (\pi_\phi^2+(\nabla\phi)^2+m^2\phi^2)-G\int d\mathbf r\int d\mathbf r'\frac{\epsilon(\bf r)\epsilon(\bf r')}{|\bf r-\bf r'|}+...\\
\end{eqnarray}
where $\pi_\phi$ is the canonical field momentum and $\epsilon(\bf r)$ is the energy density of the field. The first term represents the field action in Minkowski spacetime. The second will yield the Newtonian potential once the Newtonian limit is taken, but at this stage of the analysis represents a first-class gauge constraint, namely Gauss' law; as such it only describes gauge degrees of freedom. Finally, note that the higher-order terms indicated by ellipses include the contribution to the Hamiltonian from the transverse-traceless components of the gravitational field, which are negligible in the derivation of the interference effect (though not absent from a full description of the experiment, as we discuss shortly). It is the quantization of these omitted components that yields gravitational quanta, gravitons in this approach.

Following \cite{anastopoulos2014problems}, we extract two points from this discussion. The first is simply that the derivation of (\ref{eq:BoseH}) just rehearsed provides a justification for treating gravity as an immediate interaction between gravcats in the proposed experiments: the Newtonian model is simply what is obtained from standard quantization procedures applied to GR with scalar fields, which is itself a strategy well motivated by the success of GR and QFT. The more interesting second point, however, is that it provides us with an analysis of the nature of the degrees of freedom at play: in particular, as the direct Newtonian interaction results, in the limit, from the Gauss law gauge constraint, it is found to be only associated to a gauge degree of freedom of GR's gravitational field. In other words, assuming a `$3+1$' framework for GR, one views the Newtonian interaction as encoding no `true degree of freedom' of the gravitational field: i.e., no dynamically propagating parts of the field, as opposed to constraints. \citep[Section 6.1]{anastopoulos2021gravitational} put the idea quite starkly: ``Our analysis shows that, \textit{according to GR}, the two parts of a quantum bipartite system that interact gravitationally in the Newtonian regime \textit{do so without an intermediate degree of freedom}" (emphasis in original).

However, one must be cautious with this use of the locution `true degrees of freedom' (especially `true'!): it is a standard, well-defined notion in gauge theory, but carries significant, potentially unwarranted, ontological connotations. For instance, since indeed the Newtonian potential is a gauge constraint in $3+1$ gravity, and so not dynamically propagating, it is by definition not a true degree of freedom. But do we say on that basis that Newtonian gravity --- which after all was our best theory of the gravitational field before relativity --- was `unreal' or `unphysical'? Such a view seems implausible, though one could envision it having defenders.\footnote{In particular: although we note that dynamical degrees of freedom --- in the gauge theoretic sense --- become quantum states on quantization, constraints become operators, as we have seen.} In an abundance of such caution, to avoid conflation of a technical term with an ontological concept, we will simply avoid talk of `(true) degrees of freedom' in the following, except on a couple of occasions in which we are specifically referring back to this discussion. Indeed, in the paradigm to which we now turn, one simply does not proceed in analyzing the experiments by means of the interpretational tools used in gauge theories.

Now, as \cite[Question 5]{marletto2019answers} point out, it is not correct that the static Newtonian potential suffices for a full description of a GIE-type experiment. Our treatment focuses on the stage of the experiment in which superposed gravcats interact gravitationally while remaining (approximately) stationary. However, as we clarified in footnote \ref{fn:spin}, before that stage they must evolve into superposition, and after that stage they must evolve out of superposition. During those phases, because the packets are moving, the static field no longer suffices to describe the gravitational interaction, and graviton modes are involved. This is true even though the motion is slow (as it must be, to avoid emitting gravitons, which would destroy gravcat entanglement, as discussed in \S\ref{ssec:Belenchia}). While we are sympathetic to this point, we would also note that it is insufficient grounds for an ensuing argument on its pitch: that the Newtonian model is ill-suited, so that the GIE experiment indisputably witnesses QG (according to the tripartite analysis discussed shortly). For no one doubts that any available analysis involves some idealization of the situation; certainly this is true of the interaction stage. So the proponent of the Newtonian model paradigm could take the attitude that we should idealize the prior and posterior stages as involving no interaction at all. The issue, that is, becomes one of which idealizations to make, which is arguably a matter of theoretical perspective, which in turn is precisely what is under contention. Of course, if the experiments become sufficiently sensitive that one cannot idealize in this way, then the issue will become empirical. But in that case one would be in the regime in which one is witnessing gravitons, precisely the benchmark for positive claims of witness that proponents of the Newtonian model paradigm advocate. (Moreover, by measuring that effect one would have moved beyond the GIE experiment.)

\subsection{Tripartite models}

Notwithstanding the derivation of the non-local interaction term in the Newtonian model in a gauge theoretic approach GR, there is another view that insists on gravity, in the light of GR, as \emph{mediating} interactions between massive systems such as the gravcats: specifically, the Hamiltonian should contain no interaction term directly between gravcats, but only interactions between the gravcats and an third `gravity' subsystem. Sometimes this assumption is called `locality' since it can be secured by insisting that gravitational interactions propagate at a finite velocity. However, because considerations of causal connectibility will be relevant in the next subsection, and because the assumption could hold even if the gravcats were spatially coincident (or if effects propagated instantaneously), we will avoid that terminology. What matters here is that the only interaction terms are those defined between gravcat and gravity subsystems. 

Under these assumptions, the state of the system after splitting is in fact not (\ref{eqn:BS}), but
\begin{equation}
\label{eq:Bin}
|L\>\x|\gamma_{LL}\>\x|L\> + |L\>\x|\gamma_{LR}\>\x|R\> + |R\>\x|\gamma_{RL}\>\x|L\> + |R\>\x|\gamma_{RR}\>\x|R\>.
\end{equation}
$|\gamma_{XY}\>$ represents the gravity subsystem for a pair of gravcats located at $X$ and $Y$, respectively, so each gravitational state is that appropriate to the gravcat positions in the corresponding term. After $t$ the state is not (\ref{eq:Bapprox}) but
\begin{equation}
\label{eq:Bout}
 |L\>\x|\gamma_{LL}\>\x|L\> + |L\>\x|\gamma_{LR}\>\x|R\> +   
e^{\frac{-iGm^2t}{\delta}}|R\>\x|\gamma_{RL}\>\x|L\>+ |R\>\x|\gamma_{RR}\>\x|R\>.\footnote{And (see footnote \ref{fn:spin}) at the end of the process, after passing back through the Stern-Gerlach device the full state can be written
\begin{equation}
\label{eq:Bpost}
\Big(|X\up\>\x|Y\up\> + |X\up\>\x|Y\dn\> + e^{\frac{-iGm^2t}{\delta}}|X\dn\>\x|Y\up\> + |X\dn\>\x|Y\dn\>\Big)\x|\gamma_{XY}\>.
\end{equation}}
\end{equation}
The key thing to recognize is that in states (\ref{eq:Bin}) and (\ref{eq:Bout}) the gravity subsystem is in a \emph{superposition} of classical gravitational states. This is of course because the gravcats are in a spatial superposition and the gravity subsystem couples to the wavepackets through the Hamiltonian.\footnote{To keep things general, we will not specify a Hamiltonian, but given our assumptions it must have the following effect: there is a continuous evolution, not only of the gravcats, but also of the gravity subsystem. Moreover we assume in this experiment that time scales are long compared to $D/c$.} Thus in this model, the gravity subsystem exhibits a quantum nature, in the sense of superposing classical states.

Note that in all of the above, we have nowhere helped ourselves to the fact that GR (or, generally, knowledge that gravity is a mediator) would seem to push us into thinking of gravity as an distinct subsystem. But might there be some way to produce gravcat entanglement without treating the gravity subsystem as quantum, just by helping ourselves to such a further assumption? We already saw in \S\ref{sec:rulingoutsemiclassical} that if we treat gravity as classical in the most natural way, by invoking SCG view 2 (so taking SCG as a candidate fundamental theory of gravity) then we should expect no entanglement in the gravcat experiment. However, we will now see that this result can be generalized to the conclusion that no classical mediator can produce entanglement between quantum systems with no direct interaction.

\subsubsection{Quantum and information theoretic considerations}
\label{nonlocal}

In this section we present two results, showing from general quantum mechanical and general information theoretic considerations, respectively, that interactions with a classical mediator cannot lead to entanglement between two systems. Of course, the very setup of two local subsystems and mediating subsystems means that the tripartite paradigm is assumed from the beginning, so these results cannot settle the tripartite-Newtonian question. What they can do is show why the observation of entanglement in a GIE experiment means that the tripartite system must have a quantum mediator --- or rather, as we shall see, that the mediator not be classical. First then, a proof in quantum theoretic terms.\footnote{We thank Richard DeJong, who provided the outline of the following proof, which differs from the original proof in by \cite{MarlettoVedral} (though was inspired by it, and by a conversation with Marletto).} We will assume the simplest kind of system, in which the subsystems are 2-dimensional, but as \cite{MarlettoVedral} discuss, the extension to N-dimensional systems seems straight-forward, and thence to quantum fields in the Fock representation (although this claim has recently been challenged: \cite[\S2]{anastopoulos2022gravity}).

\begin{enumerate}

    \item Suppose that $\mathcal{H}_2$ is a 2-dimensional Hilbert space, and consider the three bit system $\mathcal{H}_2\otimes\mathcal{H}_2\otimes\mathcal{H}_2$. We make two assumptions. (a) Let the first and last slots describe qubits that interact `locally': the Hamiltonian $H=H_{12}\otimes I + I\otimes H_{23}$. And (b) Let the second slot represent a classical bit: the only observable for this subsystem is $\sigma_z$ (and the identity $I$).
    
    From (a) and (b) we have
    
    \begin{eqnarray}
        H_{12} & = & A\otimes I + B\otimes\sigma_z\\
        H_{23} & = & I\otimes C + \sigma_z\otimes D
    \end{eqnarray}
    assuming that the Hamiltonian is an observable. It follows that
    
    \begin{eqnarray}
        [H_{12}\otimes I, I\otimes H_{23}] & = & 0
    \end{eqnarray}
   As the two terms in the Hamiltonian $H$ commute, the unitary operator describing the time translations factorizes, i.e.,
    \begin{eqnarray}
    e^{-iHt} = e^{i(H_{12}\otimes I + I\otimes H_{23})t} & = & e^{-iH_{12}\otimes It}\cdot e^{-I\otimes H_{23}t}.
      \end{eqnarray}
  
   That is, since the two factors commute, we can treat the evolution as an interaction first between one of the qubits and the bit, and then between the other qubit and the bit, in either order.
    
    \item So let  the tripartite system start in a fully factorizable pure state, $\Psi$, and apply the unitary evolution:
    
    \begin{eqnarray}
        \nonumber e^{-iHt}|\Psi\> & = & e^{-iH_{12}\otimes It}\cdot e^{-I\otimes H_{23}t}|\psi_1\>\otimes|\psi_2\>\otimes|\psi_3\>\\
        & = & e^{-iH_{12}\otimes It}|\psi_1\>\otimes\sum^1_{j=0}\alpha_j|\chi_j\>\otimes|\phi_j\>,
    \end{eqnarray}
    where in the last step we have used the fact that any pure bipartite state has a Schmidt decomposition\footnote{See \cite{NielsenChuang}, section 2.5.}, in which the $|\chi_j>$ and $|\phi_j\>$ are orthonormal bases for their Hilbert spaces. Now, since the bit is classical (B), there is only one orthonormal basis for the mediating subsystem, namely $\{|0\>,|1\>\}$, respectively, the -1 and +1 eigenstates of $\sigma_z$. Hence we can continue 
    
    \begin{equation}\label{eq:schmidt}
                = e^{-iH_{12}\otimes It}|\psi_1\>\otimes\sum^1_{j=0}\alpha_j|j\>\otimes|\phi_j\>.
    \end{equation}
    
    \item\label{it:super} Since (B) the only observables on the bit are (multiples of) the identity and $\sigma_z$, there is a superselection rule: \emph{all} observables on the system commute with $I\otimes\sigma_z\otimes I$. In particular, restricting attention to the bipartite 2-3 subsystem, the most general form of an observable is $I\otimes A+\sigma_z\otimes B$, and it is straightforward to verify that this commutes with $\sigma_z\otimes I$. As a result (which can also be readily checked), the pure state $\sum^1_{j=0}\alpha_j|j\>\otimes|\phi_j\>$ is indistinguishable from the mixture $\sum^1_{j=0}|\alpha_j|^2|j\>\<j|\otimes|\phi_j\>\<\phi_j|$: the expectation values of the states agree for all observables of the general form.
    
    Therefore we can replace the pure state in (\ref{eq:schmidt}) with a physically equivalent density matrix,
    
     \begin{equation}
                |\psi_1\>\otimes\sum^1_{j=0}\alpha_j|j\>\otimes|\phi_j\>\rightarrow|\psi_1\>\<\psi_1|\otimes\sum^1_{j=0}|\alpha_j|^2|j\>\<j|\otimes|\phi_j\>\<\phi_j|.
    \end{equation}
    
    This state is separable between 2 and 3 (and indeed between 1): in general, any density matrix of the form $\sum_j|c_j|^2\rho^a_j\otimes\rho^b_j$ can be expressed as a mixture of factorizable pure states. In short, by rewriting the state in this way, we have made manifest the absence of entanglement between 2 and 3, ultimately \emph{a consequence of the classical nature of the bit}.
    
    \item Now, it should be clear that since 3 and 2 are not entangled, then -- even if 2 were quantum -- an interaction between 1 and 2 alone cannot not entangle 1 and 3! (Of course, if 2 and 3 were entangled, that entanglement could be `exchanged' between 2 and 1.) But for completeness we will show this to be the case. So consider the interaction between 1 and 2, producing the state
    
            \begin{eqnarray}
                \nonumber & \ & e^{-iH_{12}\otimes It}|\psi_1\>\<\psi_1|\otimes\sum^1_{j=0}|\alpha_j|^2|j\>\<j|\otimes|\phi_j\>\<\phi_j|e^{+iH_{12}\otimes It}\\
                & = & \sum^1_{j=0}|\alpha_j|^2e^{-iH_{12}t}|\psi_1\>\<\psi_1|\otimes|j\>\<j|e^{+iH_{12}t}\otimes|\phi_j\>\<\phi_j|.
            \end{eqnarray}
    This state is a mixture of pure states in which the 1-2 subsystem factorizes with the 3 subsystem; hence it is separable between the two qubits. (Of course, by appeal to the classicality of the bit, we can as before show that it is also separable between the 1 and 2 subsystems.)
    
    Thus we have shown using the standard assumptions of QM, the locality of the qubits (a), that no entanglement can arise between the qubits, given the classicality of the bit as expressed by (b). The argument generalizes to higher dimensional $\mathcal{H}$, and (as MV note) to fields (given a Fock representation), and hence to the GIE experiment. $\blacksquare$
    
    \end{enumerate}
    
Now, as we noted in step 3, there is a superselection rule, so that the bit state is always a mixture of $\sigma_z=\pm1$ states. But (since we assume the Hamiltonian to be an observable), the superselection rule implies a selection rule. That is,
    
    \begin{eqnarray}
        [I\otimes\sigma_z\otimes I, H] & = & 0\\
        \Rightarrow\quad [I\otimes\sigma_z\otimes I, e^{iHt}] & = & 0,
    \end{eqnarray}
so that the value of the bit is a constant. Thus not only is the bit a mixture of +1 and -1 $\sigma_z$ states, it is a \emph{constant} mixture. In other words, in this framework, the qubits have no effect on the bit, although the bit state does affect the final states of the qubits. The one-way nature of the interaction seems to refute the claim that the bit mediates any interaction \emph{between} the qubits, after all -- so it is no surprise that no entanglement arises between them! 
    
Of course, if entanglement occurs between local systems, as is expected in the GIE experiment, then the mediator is not classical, (b) or its equivalent for the system in question fails, and there are no selection or superselection rules; the mediator is (typically) neither mixed nor frozen. Moreover, in SCG the evolution is not only quantum, since the M\o ller-Rosenfeld equation governs the evolution of the mediating gravitational field, so again the field will not be frozen (though it will not be in a quantum superposition). 
    
Still, this situation makes one wonder whether the quantum framework is actually apt for representing a classical bit in the first place. Perhaps the conclusion that the mediator cannot be classical depends on adopting a hobbled description for it. One would like to see the same result in a broader setting that doesn't already suppose the quantum framework. Moreover, the proof assumes in Step 1 that the Hamiltonian is an observable, and one might somehow question that assumption. For these and other reasons it is worth sketching an alternative, more general information theoretic result.

\cite{marletto2020witnessing} uses the general information theoretic framework of `constructor theory' \citep{deutsch2015constructor} to prove that if interactions with a mediating subsystem produce entanglement between mutually isolated systems, then the mediator cannot be classical. Both the exact statement of the theorem and its proof rely on some new and unfamiliar machinery, so we will just sketch some central ideas to give a feeling for the result; for details the reader is referred to the original papers.\footnote{A theorem with analogous conclusions has also been given in the framework of generalized probability theory (\cite{galley2020no}). We focus on the constructor theory result for definiteness, but stress that the availability of similar results in other frameworks indicates the generality of the underlying argument.}

Constructor theory works at a very high level of generality, in which various subsystems are posited, each with some set of mutually exclusive states $s_i$, thought of as providing a full description of the subsystem. In addition, state transitions for individual and composite systems --- `tasks' --- $[s_i\rightarrow s_j]$ are specified as possible or impossible. For instance, suppose we have a subsystem $S$ with states $\{0,1,y,z\}$ (the numerals and letters should be understood here simply as state labels). If we think extensionally, then sets of states correspond to determinables (or `variables' in constructor theory) and individual states to their determinate values (`attributes').\footnote{Since in general one wants to allow for degenerate states, properly speaking attributes are sets of states, and variables are sets of disjoint attributes. In our examples we will simplify the notation by ignoring this extra structure.} Let us suppose then that $B=\{0,1\}$ (for `bit') is one variable, with attributes 0 and 1 (so that $B$ is for `bit'); and that $X=\{y,z\}$ is another.

Simplifying considerably (but hopefully harmlessly for present purposes), let us say that a variable $\{x_1,x_2,\dots,x_n\}$ of a system is an `observable' when there is a second system --- the measuring apparatus --- with states $\{s_r,s_1,\dots,s_n\}$, and a possible task that for any $1\leq i\leq n$ has the effect $[(x_i,s_r)\rightarrow(x_i,s_i)]$. That is, $s_r$ is the apparatus ready state, the $s_i$ are its various possible meter readings, correlated with states of the subsystem, so that the task amounts to a non-destructive measurement of the variable (the state of the measured subsystem is unchanged).

To continue our example, let us thus further suppose that $B$ and $X$ are both observables. In that case we have a further choice: their union may or may not also be an observable --- in describing a system, we choose whether or not the necessary process is possible. If $\{0,1,y,z\}$ is an observable, then $S$ is simply a classical system, with a 4-dimensional state space. But if it is not an observable, then the situation is like that in a qubit, in which $B$ and $X$ are incompatible observables, $\sigma_z$ and $\sigma_x$ say, which cannot be observed simultaneously. Broadly speaking then, this situation abstracts the notion of `complementarity' without assuming the quantum formalism. So we see that constructor theory can accommodate classical and quantum on an equal footing; the general framework of states and tasks is neutral between them, and the difference arises only from which tasks are possible. (Of course, quantum is also distinguished from classical by the capacity for entanglement, beyond any classical statistical correlation, and this distinction too can be represented.) Moreover, the generality of the framework allows for possibilities between the fully classical and the fully quantum (and indeed possibilities orthogonal to that distinction); in particular, the generalizations of complementarity and entanglement, while clearly non-classical, do not by themselves entail the full quantum formalism.

Suppose then two mutually isolated qubits (or `superinformation media', as they are represented in constructor theory), which can each interact with a mediating subsystem: of course, this state of affairs is cashed out in terms of which two subsystem tasks are and are not possible. Then, Marletto and Vedral show, if initially separable qubits end up entangled, then the mediator must be non-classical, possessing complementary variables, and entangling with the qubits, in their generalized senses.\footnote{We note that the theorem requires the union of the complementary variables not only to be unobservable in the sense we gave, but to not even be measureable by a destructive measurement that changes the state of the mediator. Nor is it required that both the variables be observables; one of them could similarly be `hidden'; in other words, since it is complementary to the other variable, it could be a non-classical hidden variable.} Thus, given the tripartite model, we again conclude that a positive result in the GIE experiment would witness non-classical behavior of gravity. In this framework we don't shoehorn a classical system into QM, and there is no requirement that the mediator be static if classical. 

Note that strictly the theorem shows that mediating entanglement requires a non-classical mediator, not that it requires a quantum one: indeed non-classical, non-quantum models accounting for entanglement mediation exist (\cite{marletto2019answers} refers to \cite{hall2018two}). In this regard Marletto and Vedral reasonably compare their result with Bell's theorem: in that case too, experiment can only show that a system is non-classical, not that it is quantum. 

Finally, now that we have seen how entanglement mediation rules out classicality, we should note that GIE experiments also bear on gravitational collapse theories (often associated with \cite{diosi1989models, penrose1994non}). Such proposals fall in the tripartite camp, but claim that the gravitational field `abhors' a superposition, and so induces a collapse whenever it deviates from a classical state (by a prescribed amount), and thus is essentially always classical. Here the implication of observing entanglement is more equivocal then for SCG: As \cite{bose2017spin} put it, such theories imply `the breakdown of quantum mechanics itself at scales macroscopic enough to produce \emph{prominent} gravitational effects' (1, our emphasis). The question of course is what counts as `prominent'. On the one hand, by Penrose's estimates, the proposed experiment, with gravcats of $10^{-14}$kg separated by $100\mu$m, the gravitational collapse time should be of the order of a second, which would be fast enough for the classicality of the field to affect any observed entanglement. And so it seems it is a `prominent' effect, so the quantum state collapses, and no entanglement will be observed. However, on the other hand, should entanglement be observed, the theories do have a tunable parameter, which could be set to prevent collapse in the currently envisioned GIE experiments, although they would place a new bound on them. But so doing is to accept that the experiment witnesses a quantum superposition of the gravitational field, which is at least against the spirit of Penrose's position, and quite possibly falls afoul of the very arguments by which he motivates it.

\subsubsection{The quantum description of gravity}\label{sec:Bose}

In this subsection we elaborate further on the quantum description of the `gravity' subsystem in these experiments, within the tripartite view.

Now, if one wishes to model the experiment according to the tripartite models paradigm, which explicitly views gravity as a physical subsystem on a par with the gravcats, then it is of course natural to start with GR: our current best physical theory of gravity after all does theorize gravity as a dynamical system with a non-trivial vacuum sector. So, within the tripartite models paradigm, GR itself would seem to justify the assumption in (\ref{eq:Bin}) that we have a tripartite system, something that seems ad hoc in Newtonian gravity proper. And indeed, this is the starting point that theorists have generally taken, in a couple of different ways.
 
For instance, the Supplementary Materials to \cite{bose2017spin} derives the phases of the various terms in the gravcat superposition by quantizing linearized GR: Briefly, one again starts by representing the classical GR gravitational field as linear perturbations in a 3+1 Minkowski background, using the stationary approximation for the gravcats, but in a choice of gauge in which only the $h_{00}$ components of the field are relevant, and then quantizes them.\footnote{Attentive readers will note that the order of operations here is (partly) opposite to the approach to quantization taken by \cite{anastopoulos2014problems}, presented in a previous section. From a rigorous point of view, it is indeed a good question how much the operations of quantization and gauge fixing commute. (For the framework of geometric quantization, commutation has been conjectured --- known under the `Guillemin-Sternberg conjecture' --- and proven in particular cases. See \cite{nlab2019}.) Our sense is that, likely, little that is relevant to our concerns hangs on the fact that the different parties involved take constraints and quantize in different orders.} The experiment can then be modeled in terms of an interaction between localized mass excitations, mediated by coherent states of the modes of the $h_{00}$ field.\footnote{As discussed in \cite{huggett2020out}, coherent states are often taken to represent classical states of quantum fields. Note that what are usually called gravitons are associated with other components of the $h$ field.} The appropriate Hamiltonian has been solved, and introduces the same phases as predicted by the Newtonian potential.

Another approach is taken by \cite{christodoulou2019possibility}, who model the experiment using quantum superpositions of the metric field, locating the experimental effect as a consequence of superpositions of gravitational redshifts (thus arguably also highlighting the role of general covariance in GR). Of course, from a quantum field theory perspective these are presumably superpositions of coherent states of the perturbatively quantized (linearized) gravitational field, but in their presentation they are simply treated as quantum states labeled by classical geometries. Starting classically, again assuming that the gravcats are heavy enough, and the duration short enough that they remain (approximately) stationary during the experiment, the metric field due to a body of mass $m$ and radius $R$ can be approximated by

\begin{equation}
\label{eq:metric}
    ds^2=(1-2\phi(\vec x))dt^2-d\vec x^2.
\end{equation}
Outside the body, $\phi(\vec x) = Gm/r$ is just the Newtonian potential; while inside the body one takes $\phi(\vec x)=Gm/R$, a constant. (Compared with [\ref{eq:WF}] we now set $c=1$ and consider the interior as well as exterior field.) As always, we are in the regime of linearized gravity, so that for a pair of gravcats, the classical geometry will be the sum of two such solutions. It is then straight-forward to calculate the proper time elapsed along the worldline of (a point in) one gravcat at a distance $\delta\gg R$ from another:

\begin{equation}
\label{eq:CRRS}
    \tau = \int_0^T ds = \int_0^T \sqrt{1-2Gm/R-Gm/\delta}\ dt\approx T\cdot(1-Gm/R-Gm/\delta),
\end{equation}
showing the usual GR time dilation effect of a nearby mass. If $d\gg R$, most of the effect will come from the mass of the gravcat itself, but we also see that there is a contribution dependent on the distance to the other gravcat.

Thus, the components of the superposition (\ref{eqn:BS}), corresponding to different gravcat separations, correspond to different metrics, and hence different proper times. Now assuming that the (classical) gravitational field labels what is actually some coherent state in a quantum treatment of gravity, which can superpose, the full state is therefore (\ref{eq:Bin}), with $|\gamma_{LR}\rangle$ etc describing different states of the metric field of GR.  And thus the different terms of the superposition are associated with different proper times and will develop relative phases (relative to a common lab time, $T$). In particular, a gravcat pair has a phase $e^{-im\tau}$ from its mass energy, or (ignoring a common phase) $e^{-iGm^2T/\delta}$ from (\ref{eq:CRRS}) --- the very phase one expects (\ref{eq:Bapprox}) from the Newtonian potential, but now seen to be a relativistic redshift effect. Thus the very same relative phase between the components of the gravcat superposition, and ultimately the very same entanglement, is predicted by treating metric states as superposable quantum states (to leading order). 

Both of these treatments, which in different ways `quantize' the metric field of GR, treat gravity as a physical object, which naturally should be treated as a third subsystem in addition to the gravcats, and which \emph{then} must be quantum (or at least non-classical) by the general quantum and information theoretic arguments, if the gravcats become entangled as expected. Thus, from this point of view, the derivation of the effect by appeal to the Newtonian potential with which we presented it on the naive first pass is an approximate heuristic. (Recall for contrast that the analysis by Anastopoulos and Hu in \S\ref{sec:paradigms} --- or rather their interpretation thereof in favor of the Newtonian model view --- explicitly renders the Newtonian potential as uninformative of the physical parts of the gravitational field!)

A word of caution: because modeling the system in this way involved an appeal to GR, it is tempting to think that the finite speed of propagation of the gravitational interaction in GR is necessary and sufficient for the system to be tripartite. But while a finite propagation speed seems to require that the system be tripartite, we would nonetheless now like to stress that it is by no means necessary. This point can be seen clearly by adopting a geometrized formulation of Newtonian gravity: Newton-Cartan theory.\footnote{There is a robust philosophical tradition of moving to the geometrized Newton-Cartan theory, in order to more sharply compare the features of Newtonian gravitation with the features specific to its relativistic successor theory: e.g. \cite[ch. 4]{malament2012topics},\cite{malament1995newtonian},\cite{weatherall2014singularity}.} In fact, as we demonstrate in the appendix, whereas the gravitational redshift effect noted in \cite{christodoulou2019possibility} is specific to their relativistic setting, the surrounding argument --- namely, that a superposition of spacetime geometries demonstrates the quantum nature of gravity --- is not. In particular, one can perform an independently motivated analysis of the gravcat experiment in the framework of Newton-Cartan theory, which is analogous to their relativistic analysis: in both, gravcat pairs develop different phase factors with respect to different, superposed, spacetime geometries, leading to entanglement. Again, all that is important is the decision to treat gravity as a mediating subsystem, a decision whose subsequent execution proceeds analogously across both theoretical contexts.

Moreover, this example emphasizes that the mediating subsystem need not be a dynamically propagating field for the entanglement theorems to apply, and hence for the GIE experiment to witness quantum gravity. (Note that there exists a quantum treatment of gravity formulated according to Newton-Cartan theory \citep{christian1997exactly}.) In other words, even though the theorems entail that a successful GIE experiment witnesses the non-classical nature of gravity, it does not, without further assumptions, witness perturbatively quantized linearized gravity. Of course, it is possible that these further assumptions are supplied by the concurrent commitment to GR in the background of an analysis of the Newtonian-regime experiment. This is a point we will discuss in \S\ref{ssec:Belenchia}, in the context of a thought experiment that has been analyzed by \cite{belenchia2018quantum}. (A related point --- that the GIE experiments may be interpreted as a degenerate case of a local, retarded GIE protocol where corrections due to retardation/causality considerations are negligible --- will be discussed shortly.)

\subsection{Why `paradigms'?}

At this point, we should justify our description of the Newtonian and tripartite (kind of) models not just as models, but as \emph{paradigms}. First, by this term we do not mean to bundle together the multiple senses and aspects described in \cite{thomas1962structure}. Rather, we specifically mean that the models reflect choices of theoretical standpoints between which rational argument alone cannot adjudicate. Importantly, both models are motivated from the joint methodological principle that the theory of Newtonian gravity, as relevant in the naive description of the GIE experimental setup, should be interpreted from the viewpoint of our best theory of gravity --- where all recognize this theory to be GR. Rational indeterminacy arose because each model picks out different aspects of GR as central: in one case it is the lesson that gravity is non-instantly propagated, in the other case the nature of the Newtonian limit. And while proponents of both models can be brought to \emph{understand} the analysis of the other, they will fail to see the complete relevance it has for the other --- the degrees of freedom analysis on when to call putative gravitational systems physical is simply mutually exclusive from viewing the gravitational field as propagating and vice versa. Hopefully the preceding discussion makes clear these competing theoretical stances, why they are both reasonable, yet incompatible: exactly the situation in which we are willing to speak of `paradigms'.

One might still worry that there is some argument that we have missed, relying on deeper principles agreed to by the partisans of both paradigms, which would after all rationally require one group to abandon their commitments, leading to the collapse of our talk of paradigms. Presentations of this work\footnote{For instance to audiences at the Center for Philosophy of Science at the University of Pittsburgh, and the Rotman Institute at Western University.} have indeed encountered such resistance to calling the positions `paradigms' in our sense, some rehearsing considerations already discussed, and some raising new arguments to be considered here. It is worth noting, however, that while those arguing against paradigm talk have done so on the basis that one of the models is \emph{clearly} the right one, they have not agreed about which one in fact is the right one. So this datum is at least compatible with the thesis that our interlocutors are in fact adopting different paradigms in our sense! Nonetheless, we do take their arguments seriously, and indeed addressing them will help clarify our claim.

\textit{Objection:} By denying gravity the status of a causal field, the Newtonian model is committed to such an unreasonable refusal of physical background knowledge that it is disqualified as sensible physical theorizing by any paradigm-independent standard of modern physics.\\
\textit{Reply:} The Newtonian model does take into account physical background knowledge --- arguably, just the same knowledge (it just weighs the relevance of the analysis given in \S\ref{sec:Newtonianmodeling} completely differently). Indeed, that the Newtonian model takes into account the same physical background knowledge is emphasized, ironically, in a recent article by \cite{christodoulou2022locally}. In the article, which is intended to reinforce that GIE experiments indicate quantum superposition of spacetimes, the authors distinguish a `slow-motion' approximation (sources moving at non-relativistic speeds), where the gravitational interaction is still local (i.e. causal), and a `near-field approximation' (sources are much closer together than time elapsed through the experiment, in natural units). The `Newtonian limit' that reproduces the naive model, claim the authors, denotes the overlapping regime where both approximations are satisfied. Now, as they show, in the slow-motion, not near-field approximation regime, locality considerations show up as corrections between the employ of laboratory time function (as in the Newtonian limit) and a retarded time function. But, setting up retarded GIE experiments in the slow-motion, not near-field regime is ``not realistic for the foreseeable future'' and seeing the effects of slow-motion correction terms to the Newtonian limit analysis of the GIE experiments is ``unlikely reachable'' (p. 4). Thus, although there are retarded GIE protocols that they point to on a further horizon of experimentation (p. 5) that, given physical background knowledge, would preclude a Newtonian model description, the GIE experiments under scrutiny here do not: precisely because they are in the Newtonian limit regime of our fundamental gravitational physics.\footnote{Note the main point of their article is ultimately to develop the physical picture of superposition of spacetimes in GIE experiments: ``The physical picture arising from the analysis is that information travels in the quantum superposition of field wavefronts: the mechanism that propagates the quantum information with the speed of light is a quantum superposition of macroscopically distinct dynamical field configurations'' (p. 5).}

To appreciate further that proponents of both paradigms deal with the same kind of background knowledge --- but simply weigh it differently --- it is perhaps helpful to draw a connection to a similar case: the EFT/renormalizability paradigm distinction regarding QFTs. In the traditional renormalizability paradigm, non-renormalizable theories as such are seen as defective; renormalizable theories, being well-defined, thereby vindicate non-renormalizable theories derived from them in low-energy limits. In contrast, in the EFT paradigm, a theory is seen as effectively independent from underlying theories, and renormalizability drops out as a criterion. Now, both views acknowledge a joint background knowledge of decoupling theorems: statements about how the physics of one theory in the QFT framework is effectively decoupled from the other. Nevertheless, it is only the EFT view which sufficiently weighs these kind of arguments for re-thinking what we mean by a well-defined theory: that based on such decoupling theorems (1) non-renormalizable theories can count as effectively renormalizable, and (2) renormalizable theories are replaceable by non-renormalizable theories for any regime of validity. As a consequence, the confirmation of theory A through testing a theory B that reduces to A will be significantly higher in the renormalizability paradigm then in the EFT paradigm.

\textit{Objection:} \emph{Both} paradigms take GR as their starting point, thus accepting that gravity is well-described by a dynamical classical field in the appropriate limit. What then is the state of this field supposed to be if we have superposed gravcats, and their consequent observable entanglement? The general information theoretic theorems show that it cannot be a classical state if the interaction is mediated by gravity in the appropriate sense, and no other mechanism in which gravity remains in a classical field state has been proposed (if any be possible). Doesn't then entanglement require a quantum superposition according to the tenets of both paradigms?\footnote{This objection is our rendering of a point put to us forcefully and patiently by Carlo Rovelli.}\\
\textit{Reply:} First, one might accept GR in its regime, but not accept that gravity is a `field' (broadly speaking), classical or quantum, more fundamentally. One might have a specific alternative theory in mind (gravity as ancilla, perhaps), but more likely, one might simply be neutral on the topic. That would be an unusual stance for a theorist, but would be less implausible for an experimentalist, whose goal of probing the quantum + gravity regime requires only very broad theoretical commitments, as opposed to a more specific package of causal lore. Either way one could resist the objection, because one is not committed to any field state at all.

One might question whether the derivation of the Newtonian model in \S\ref{sec:Newtonianmodeling} is undermined by this position. After all, one can certainly motivate such a derivation as an implementation of LEQG. However, as we primed the reader in \S\ref{sec:prelude} one could also motivate it simply as a minimal way to combine central insights from a conflicted corpus that includes GR and QFT.

Most theorists, however, seem to accept the EFT approach to QG, and thus that `really' gravity is a field, and so a third subsystem as conceived in the tripartite approach. Isn't such a person thus bound to accept that paradigm, and conclude that gravcat entanglement would witness a non-classical state of the field? This thought leads to the second reply: `yes and no'. The `really' provides important wiggle room, because it could mean '(more) fundamentally'; in which case yes, in a more complete physical description, gravity is in a superposition, but also perhaps no, in the model \emph{most apt} for the experiment, gravity is not represented as quantum, and hence such behavior is not witnessed. That is, the dichotomy between Newtonian and tripartite paradigms need not be one concerning the fundamental nature of gravity, but one concerned with the best model of a specific situation, and hence where one sets the bar for experimental observation of quantum behavior. 

Now, that one has admitted that, in a full description, gravity is necessarily quantum is surely a good reason to accept the tripartite model, which represents that fact. However, there are also reasons not to. First, one could adhere to the maxim that empirically equivalent models that include fewer details are to be preferred, on the thought that extra details are superfluous. And one could apply such a maxim to prefer the Newtonian model. Second, while rejecting the idea that only the so-called `true' degrees of freedom are physical, that the Newtonian potential merely represents a constraint surely has some import; that it would be artificial to replace it with a dynamical subsystem, perhaps. That is, the very possibility of the Newtonian model (including, remember, its derivation) shows that gravity is not playing the role of a dynamical subsystem in the experiment. Finally, the person we are considering can articulate clearly the kind of experiment in which they would agree that the quantum nature of gravity was observed, one in which no mere potential could do the job: for instance, the kind of graviton decoherence experiment that we have mentioned, as well as perhaps the retarded GIE protocols discussed by \cite{christodoulou2022locally} that were referenced above.

And yet, `really' the interaction \emph{must} be quantum (or at least non-classical). We emphasize that we do not offer these arguments as decisive. Quite the contrary, our point is that there are compelling arguments on both sides, yet the considerations are not of the kind that can be settled empirically, or by more fundamental principles. They rather `influence decisions without specifying what those decisions must be' (\cite[362]{kuhn1977objectivity}). Which is why we describe the positions as `paradigms'.

\textit{Objection:} The analysis given in \S\ref{sec:Newtonianmodeling} leading to the Newtonian model paradigm depends on a specific choice of gauge (that is, the ADM gauge). But conclusions about the quantum nature of gravity ought to be gauge-independent; put another way, the conclusion that gravity is classical depends on an arbitrary choice.\\
\textit{Reply:} There is nothing per se inappropriate about fixing a gauge to construct a model; one does it all the time. And that is exactly what is done here, \emph{as a means} to take the Newtonian limit, as one must to describe the GIE experiment.\footnote{A completely gauge-independent version of the analysis has been put forward in \cite[\S3]{anastopoulos2021gravitational}, dismissing such concerns. But note: even this more general analysis can only identify the Newtonian potential as the relevant constraint remaining in the non-relativistic limit upon restricting to the ADM gauge. We do not see this as a bug, but rather as a feature of the analysis.} (In line with the previous reply, one can of course agree that `really' the field is quantum, yet it is most appropriate to construct a gauge fixed model with gauge-relative observables.)

\section{Witnessing gravitational quanta?}\label{ssec:Belenchia}

Instead of a practical experiment (though see the end of the section) \cite{belenchia2018quantum} have proposed a thought experiment, in which gravity putatively has to be treated as quantized mediator in order to avoid a paradox. As that (thought) experiment has been taken as a point to favor actually modeling the GIE experiment from a tripartite paradigm (for instance, parts of the introduction of \cite{belenchia2018quantum} can be interpreted to this effect), we clarify why it indeed has no role in removing the clash between bipartite and tripartite proponent. (In fact, we take this analysis as a nice display of how our previously introduced distinction between the two modeling paradigms can do real work.)

Let us first critically rehearse the proposal behind \cite{belenchia2018quantum}. (In line with all previous instances of quantum gravity experiments, we shall discuss this (thought) experiment in a self-contained manner as well.) To start, consider two observers, Alice and Bob, a distance $D$ apart. Assume that Alice has already used a Stern-Gerlach apparatus oriented along the $z$-axis to prepare a massive particle with positive $x$-spin in a state
\begin{equation}
    \label{eq:Ast1}
    \frac{1}{\sqrt{2}} (\ket{L}_A \ket{\uparrow}_A + \ket{R}_A \ket{\downarrow}_A),
\end{equation}
where $L, R$ denote the left and right of Alice, a distance $\delta\ll D$ apart. (Assume further that the state preparation was adiabatic: slow enough that it does not become entangled with relevant fields by emitting radiation.) Now suppose that Bob has placed a massive particle into a trap, such that the particle effectively registers no effect of outside systems. In these circumstances, were Alice to pass her system (again adiabatically) through a ``reversed" Stern-Gerlach apparatus, recombining left and right packets of (\ref{eq:Ast1}) at a single location, then since the particle has remained free of entanglement with other systems, the two spin states would interfere, producing a final state of positive $x$-spin. Let us suppose that she indeed completes this task in a time $T_A>0$ in the lab frame.

\begin{figure}[h]
    \centering
    \includegraphics[width=0.4\textwidth]{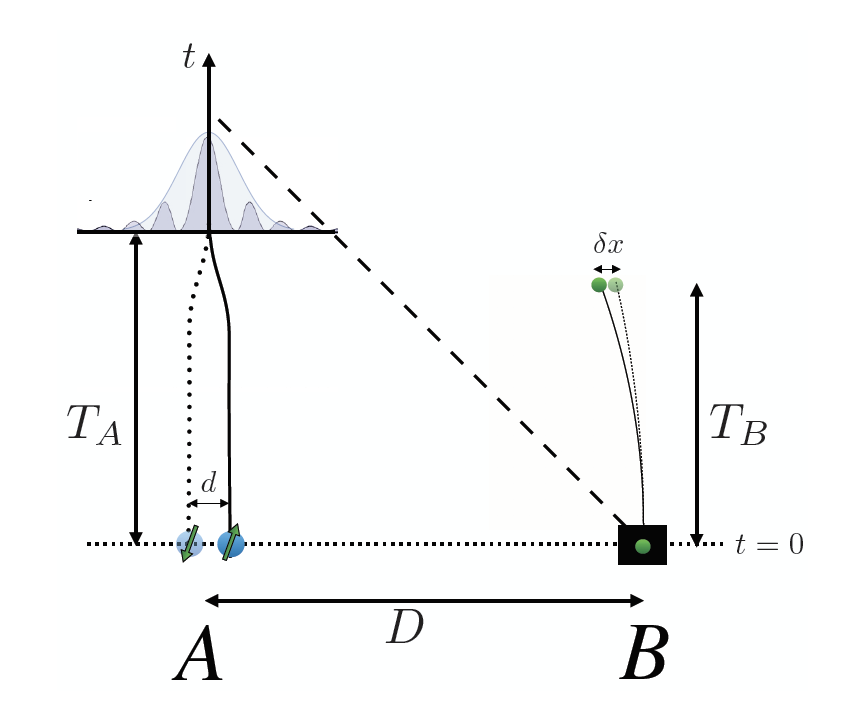}
    \caption{Arrangement of the Gedankenexperiment of \cite{belenchia2018quantum}, (figure 1).}
    \label{fig:mesh1}
\end{figure}

But consider what happens if instead, for a time $T_B>0$, Bob lets his particle leave the trap and thereby experience the gravitational field of Alice's particle. Because the two packets are at different distances $D$ and $D-\delta$ from Bob's particle, associated with different gravitational forces on Bob's particle, the two particles will entangle: the right (left) packet of Alice's particle becomes correlated with a packet of Bob's slightly more (less) deflected to the left. There then \emph{seems} to arise a clash between `complementarity' and causality, if Alice and Bob are spacelike separated (i.e., $T_A, T_B < D$, since $c=1$). It seems, that is, that the entanglement of Alice and Bob's particles caused by their interaction entails that the two packets of (\ref{eq:Ast1}) will no longer interfere, and Alice would no longer find a positive $x$-spin state after passing her system through the ``reversed" Stern-Gerlach apparatus.\footnote{Note that such an observation will be statistical in nature: that the final state is positive $x$-spin up requires observing all spin up outcomes in an ensemble of particles prepared by multiple runs of the experiment. Now, even unreliable, probabilistic signaling would violate causality, but this would still require more than a single run of the experiment: a single positive $x$-spin measurement gives no information about whether Alice's particle is in such a state or in the mixture of $\ket{\uparrow}$ and $\ket{\downarrow}$ expected if it entangles with Bob's particle. To maintain the strictest constraints on causality, Alice and Bob should remain spacelike for the whole series of runs; but if we put to one side the possibility of past runs influencing the outcomes of future runs, we only require that Alice and Bob remain spacelike during each run, as assumed here.} But this would mean that Bob's release of his particle from the trap --- which only lasted $T_B<D$ --- would have had a faster-than-light effect on Alice's system. Apparently, by deciding whether or not to release his particle, Bob could transmit a bit of data to Alice superluminally, violating causality. If, conversely, Alice's system still did exhibit interference, observed as a final $x$-spin state, then we could conclude that it remained in a superposition (\ref{eq:Ast1}) of \emph{both} left and right packets during the experiment; yet Bob could measure the deflection of his particle to determine which \emph{one} of the two paths Alice's particle actually took --- violating `complementarity'!

Fortunately (for orthodoxy), as \cite{belenchia2018quantum} show, the apparent clash can be dissolved if the following two more subtle requirements are integrated into the modeling:

\begin{description}
\item[Minimum length requirement (MLR):] distances in a QG context can only be resolved up to some `Planck length' $L_p$. Belenchia et al argue for this by appealing to the (dominant term of the) vacuum fluctuations of the Riemannian curvature, averaged in a spacetime region of radius $R$: its correlation function of the (linearized) tensor has a dominant term of form $L_P/R^3$; integrating the geodesic equation over time $R$ yields an estimated fluctuation in the relative position of two bodies of form $\delta x \sim L_P$. However, note that there is not only one possible argument that Planck length is a minimal distance in QG (as discussed further below), and that the particular value of the Planck length does not do much work in any case.

This requirement can be coupled with the fact that the spatial displacement of Alice's wavepackets leads to an effective (static) quadrupole moment (the dominant gravitational effect since gravity features no dipole radiation). Since the gravitational field difference associated to the effective gravitational quadrupole $Q_A$ is $Q_A/D^4$, using $m_B \ddot{x} = F_\text{Newton}$ it can be estimated that Bob's particle of mass $m_B$ will have been displaced by an amount $\frac{Q_A}{D^4} T_B^2$ after time $T_B$.\footnote{Using the integration estimate $\ddot{x} \sim \delta x T_B^2$.} But MLR means that this distance has to be greater than $L_P$ for Bob to resolve the path of Alice's particle: that is, it follows that for Bob to obtain `which-path' information about Alice's particle, we must have $\frac{Q_A}{D^4} T_B^2 > L_P$, or $Q_A T^2_B>D^4$ in $L_P=1$ units --- what we dub the MLR --- fulfilled.

\item[Quantized radiation requirement (QRR):] when Alice recombines the wavepackets of her particle during time $T_A$, gravitational radiation is emitted in quanta of frequency $f\sim 1/T_A$. Now, the total energy radiated is given as $E \sim \left( \frac{Q_A}{T_A^3} \right)^2 T_A$.\footnote{See \cite{rovenchak2018radiation}, section 3, for a derivation of this quadrupole formula. Note that this radiation is in the form of physical, transverse gravitons. In contrast, the quadrupole field relevant to the MLR is due to the gauge, Gaussian part of the field.} Given the standard de Broglie energy-frequency relation $E = Nf$, it follows that the number of quanta emitted will be $N \sim (\frac{Q_A}{T_A^2})^2$. However, if \emph{any} quanta at all are emitted, then Alice's particle will become entangled with the gravitational field, and coherence will again be lost: so we must have $N<1$ if she observes interference. In other words, the requirement $Q_A<T^2_A$ --- the QRR --- needs to be fulfilled if Alice's particle is to remain in a pure state. (This inequality is so dubbed because of the derivation presented here, but note that later we will ask whether the QRR could be derived without the assumption of quantized gravitational radiation.)\footnote{In principle this radiation could also be measured by Bob at a lightlike separation for its emission, further entangling with him. However, he could not also obtain which-path information, since that would collapse Alice's particle into a single packet, preventing recombination, the source of this radiation.}
\end{description}

If the localization and radiation requirements are appropriately taken into account, it can easily be shown that there is indeed no paradox, i.e. no clash between causality (no superluminal signalling) and complementarity. Consider two cases:

\begin{description}
    \item[$Q_A < T_A^2$:] The QRR is explicitly fulfilled so that Alice can recohere her packets, but the MLR is not fulfilled and Bob cannot measure the path taken by Alice's particle: in order for Alice and Bob's procedures to be spacelike $T_A,T_B<D$, so $T^2_AT^2_B<D^4$, and in the current case $Q_AT^2_B<D^4$, contrary to the MLR. Since it is impossible for both Alice and Bob to perform their procedures, causality and complementarity will not clash.
    \item[$T_A^2 < Q_A$:] The QRR is explicitly violated so that Alice's particle is entangled with the gravitational radiation emitted when Alice attempts to recohere her particle, thwarting that attempt. On the other hand, for a suitably large value of $Q_A$ it is perfectly possible for the MLR to be satisfied,
    \footnote{\label{footnotecondition} Concretely, for the MLR to be fulfilled we need to have $Q_A T_B^2/D^4 > 1$. Given that $0 < T_B^2/D^2 < 1$, the MLR can only be fulfilled if and only if $Q_A > D^2$ (notably $D^2 > T_A^2$). So, actual fulfillment of MLR requires $Q_A$ to be larger than $D^2$ (and not just larger than $T_A^2$).} Again, since we do not have both which-path information, and recoherence, there is no clash of causality and complementarity.
\end{description}

To sum up, successful completion of the experiment requires that both MLR and QRR are satisfied, but as the analysis of \cite{belenchia2018quantum} shows, on the assumption of perturbatively quantized linearized gravity --- which we will call the `graviton hypothesis' --- they cannot be simultaneously satisfied. But if the successful completion of the experiment is thus ruled out, then no clash between causality and complementarity arises.  However, the analysis in fact only shows that the graviton hypothesis is sufficient to avoid the clash of causality and complementarity, not that it is necessary (as some of their statements suggest).\footnote{\cite{avila2022quantum} argue that the assumption of quanta is not even sufficient to evade paradox: under the assumption that there is not one particle trap on Bob's side but a sufficiently high number $N$ of particles traps, one can obtain decoherence arbitrarily fast by releasing all particles at once --- even if the release of a single particle would not lead to decoherence. We leave aside whether their modification leads to some distinctively new paradox that would require further strategies to resolve.} But could the case for MLR and QRR really be made without appeal to gravitons? Clearly, this question (in particular with regard to QRR which concerns transverse, i.e. dynamically propagating, gravitons) is important in evaluating the role this thought experiment can play in defending a claim that the gravcat experiments supply a tabletop witness of the quantum nature of gravity.

\begin{description}
\item{On MLR:} The derivation of the MLR à la Belenchia et al. operates via a minimal length argument that adheres to vacuum fluctuations. Notably, though, there are a number of arguments for a minimal length (see \cite{hossenfelder2013minimal}), many of which do \emph{not} make recourse to a quantization of gravity at all. Consider for instance the estimate on distance measurement limitations originally going back to \cite{salecker1997quantum} (here presented following \cite{hossenfelder2013minimal}, section 3.1.3): if the position of a non-relativistic clock is known up to $\Delta x$, then by the Heisenberg uncertainty relation its velocity can only be known up to $\Delta v = \frac{1}{2M \Delta x}$. Say the clock is a light clock, and time is measured by a photon going back and forth between two mirrors separated by the distance $D$. Within one clock period $T=2D$ ($c$ is set to 1 again), the clock thus has an uncertainty in position of $\Delta x + \frac{T}{2M\Delta x}$. The minimal value for $\Delta x$ (found from a simple variation exercise) is $\Delta x_{\text{min}} = \sqrt{T/M} = \sqrt{(2D)/M}$. Requiring a minimal operational length of the clock larger than its Schwarzschild radius (otherwise the clock would be gravitationally unstable), leads to $\Delta x_{\text{min}} > \sqrt{4G} \sim l_P$. Again, no quantization assumption about the gravitational field itself has been used to arrive at this estimate! The fact that vacuum fluctuations are not needed to establish MLR shows that there is something misleading about \citeauthor{belenchia2018quantum}'s statement that ``both vacuum fluctuations of the electromagnetic field and the quantization of electromagnetic radiation [and, analogously, in the gravitational case] were \textit{essential} for obtaining this consistency [between complementarity and causality]" (p. 5, our stress).

\item{On QRR:} As argued in footnote \ref{footnotecondition}, the condition $Q_A < T_A^2$ --- and no weaker condition such as $Q_A < T_A^2 + C$ with $C > 0$ --- has to be taken as a criterion so that Alice's particle state remains coherent to resolve the paradox. To illustrate this point: one might for instance motivate a bound from simply requiring the radiated classical quadrupole energy to be negligible, i.e. $E \sim \left( \frac{Q_A}{T_A^3} \right)^2 T_A << 1$ in order to ultimately exclude \textit{any} kind of interaction caused by classical radiation linked to Alice's closing of her quadrupole. This would entail a `radiation requirement' of form $Q_A << T_A^{5/2}$ (call this CRR) which is a weaker bound than that from QRR for $T_A > 1$ and a stronger bound for $T_A < 1$.\footnote{Concretely, $T_A^2 < T_A^{5/2}$ for $T_A > 1$, and $T_A^2 > T_A^{5/2}$ for $T_A < 1$.} The interesting case is that of $T_A < 1$: then, the CRR can be violated \textit{while} the QRR is fulfilled. So, if the violation of CRR can really be taken to entail that there is relevant interaction caused by the closure of the quadrupole and thus a decoherence effect for Alice's spin-state through any mechanism \textit{whatsoever}, the paradox would nevertheless remain: for $T_A < 1$, violation of the CRR would always go along with a violation of the MLR, as the violation of the CRR is accompanied by a fulfillment of the QRR (which in turns implies a violation of the MLR, as shown before).\footnote{More precisely, fulfillment of MLR, violation of CRR and fulfillment of QRR require (i) $Q_A T^2_B>D^4$, (ii) $Q_A > T_A^{5/2}$ and (iii) $Q_A < T_A^2$, respectively. For $T_A < 1$, (ii) and (iii) combine to $T_A^2 > Q_A > T_A^{5/2}$. But $Q_A < T_A^2$ implies---given that $T_A, T_B < D$---$Q_A T_B^2 < D^4$, i.e. the violation of MLR.}
Thus, if the only grounds for the QRR lie in the postulate of radiation quanta, there is quite a straightforward sense in which it requires gravitons, and hence yields a resolution of the paradox that is due to gravitons, as the authors claim (even if the MLR does not require the graviton hypothesis).

However, this claim has been questioned by \cite[p. 22]{anastopoulos2021gravitational}:
\begin{quote}
\dots 
a close reading of the argument [of \citeauthor{belenchia2018quantum}] shows that vacuum fluctuations need not be quantum, and that the restoration of quantum complementarity only requires a decoherence mechanism---spontaneous emission of discrete quanta being only one of possible scenarios. Hence, the arguments of [\citeauthor{belenchia2018quantum}] do not rule out theories in which gravity is treated as a classical stochastic field that causes decoherence to quantum matter, which are properties that any mathematically consistent quantum-to-classical coupling must have, anyway. 
\end{quote}
First of all, \cite[p. 6]{belenchia2018quantum} agree with the point made above that the MLR can be realised without quantisation --- and thus they should be happy to grant MLR in any case. So if one buys into the set-up of the argument as presented so far, the central issue is this: to what extent does the need for a decoherence mechanism establish a requirement like QRR (or one at least as strong as QRR for $T_A > D$). Moreover, the account of a purely classical mechanism would have to explain in what sense the naive classical account used for deriving the CRR above goes wrong. In any case, we do not know of any such explicit account, and it is difficult to envision how it would go.
\end{description}

The immediate upshot of our critical rehearsal of \cite{belenchia2018quantum} is as follows then: In the end, in order to resolve the paradox noted by \cite{belenchia2018quantum}, no explicit viable alternative has been offered to that of assuming gravitons, that is, to assuming that (i) gravitational radiation comes in (some form of) quanta, and that (ii) this `quantizing' up of gravitational radiation is aptly described in the terms provided by perturbatively quantized linearized gravity. 

But how does the \cite{belenchia2018quantum} experiment bear on our assessment of the GIE experiment? In particular, having articulated and distinguished between the Newtonian model and tripartite paradigms' analyses of the GIE experiment above, we may ask: does the new thought experiment lead the proponent of the Newtonian model to a new conclusion? Does it help the proponent of the tripartite paradigm?

The Newtonian model proponent is not moved: notably, the decisive regime for the paradox is a (special) relativistic one, whereas, for the proponent of the Newtonian model, it is a non-relativistic regime. Recall that the proponent of the Newtonian model asks which gravitational degrees of freedom (in the precise sense discussed earlier) of the (3+1) general relativistic theory still remain relevant in the fully non-relativistic regime of the GIE experiment, and answers that there are none: the interaction is mediated entirely via the Newtonian potential qua Gauss constraint. There is thus no reason for the proponent of the Newtonian model paradigm to rethink her analysis in the face of a thought experiment unconcerned with the non-relativistic regime. In short, while \cite{belenchia2018quantum} shows that the tripartite paradigm will resolve the given paradox, this does not imply that it must also be imported to the GIE experiments; our claim that they are `paradigms' stands.\footnote{We thank Flaminia Giacomini for discussions of the import of the paradox.}

If what there is to learn from the thought experiment falls silent on the proponent of the Newtonian model, what then are the stakes for the proponent of the tripartite paradigm? First, note that the tripartite paradigm proponent is not in need of learning that gravity is quantum in the relativistic regime, as she is already taking such a stance for granted --- having contented herself with the quantum nature of gravity already (by hypothesis) on show in the GIE experiment in the non-relativistic regime.\footnote{Admittedly, the theoretical argument may, however, boost her confidence in that the theoretical picture of gravity should at this scale include gravitons specifically.}

On the other hand, this is a new regime! Let us then, finally, consider how the \citeauthor{belenchia2018quantum} thought experiment could indeed lead to an experiment in the relativistic regime that would provide a QG witness in terms of gravitons --- to proponents of both paradigms (regarding the GIE). So now suppose that the experiment is carried out as described \emph{except} that Alice attempts to recombine her packets quickly enough that $T^2_A\lesssim Q_A$. The analysis indicates that she will fail because the quadrupole of the packets emits a single graviton, and so ends up in a mixed state. In this modified experiment, the absence of interference apparently demonstrates the existence of gravitons. (Conversely, if she recombines them slowly enough that $T^2_A\gtrsim Q_A$, Bob's failure to determine which path the particle took could be taken as evidence for a minimum length.) Indeed, as we mentioned in footnote \ref{ftnt:GGE}, it has been proposed that the quantum nature of gravity might be observed indirectly in much this way: a Gravitationally Induced Decoherence experiment (in contrast with GIE above).

\section{Making gravity quantum}\label{secMakingQuantum}

Let's take stock of the ground we have covered so far. We were interested in a new class of proposed experiments relevant to quantum gravity phenomenology --- GIE experiments --- in which a tabletop pair of `gravcats' promise to probe the quantum nature of the gravitational coupling between them. We were interested, that is, in articulating the punchline of  actually performing such experiments within the not-so-distant future. So: what is that punchline?

\subsection{Witness and control traditions}

We considered one possible answer, inspired by remarks of many of the actors involved within the quantum gravity phenomenology community: that these experiments would provide indirect \emph{witnesses} of gravity's ultimately quantum nature. We contextualized such a view as situated within a tradition of experiments supplying increasingly sophisticated or subtle witnesses of gravity's quantum nature --- locating the origins of such a tradition in \citeauthor{page1981indirect}'s claims in \citeyear{page1981indirect} to have provided indirect evidence of quantum gravity, given a unitary interpretation of QM. From this perspective, the newly proposed class of experiments constitutes the vanguard of that tradition, which emphasizes the achievement of ever new kinds of \emph{evidential support} for QG.

We saw, however, that, given our current best physics, the question of whether successful GIE experiments witness the quantum nature is paradigm dependent. In the tripartite paradigm, we saw exactly the sense in which the proposed experiments would provide a novel and subtle kind of witness for gravity's ultimately quantum nature: our punchline apparently identified! Yet, on the Newtonian  paradigm, the sense in which the witness provided is one that concerns the quantum nature of \emph{gravity} is muddied at best. The upshot, then, is somewhat muddled. 

We would like now to suggest that there is another possible answer, entirely, as to the punchline of performing the new experiments: simply to obtain the \emph{ability} to perform them! As will become clear, this answer makes sense of several outstanding threads in the dialectic just summarized. Most importantly, it provides a second punchline, which ultimately yields a strong motivation according to \emph{either} of the paradigms.

To begin, consider the immediate upshot of our being able to perform GIE experiments. Even from their naive presentation above, we would have controlled the gravitational coupling between two gravcats. In the tripartite paradigm, this is (further) to say that we have controlled the gravitational subsystem that mediates entanglement between the two gravcats. On the Newtonian model paradigm, by contrast, we might rather say that we have controlled the constraint part of the gravitational gauge coupling between quantum systems. Regardless of paradigm choice then, we would have achieved `control' in a new regime. This situation is comparable with demonstrations that ever bigger macromolecules can remain in superposition for a non-negligible time: despite some remaining Bohrians, near consensus has it that some form of quantum universalism holds. It's also comparable with increasingly careful Bell-violation experiments. Lastly, compare the enthusiasm for GIE experiments with that extended to COW in the history of experimentation within quantum gravity phenomenology: in a nutshell, it is \emph{doing} something not yet done with gravitating quantum material systems, or doing something that until now could not have been done in quite so careful a way with the same. Thus, we might situate the proposed experiments not only as contributing to the witness tradition (according to one paradigm), but also as contributing to the vanguard on an entirely orthogonal tradition of experimentation within quantum gravity phenomenology: a tradition of increasingly sophisticated control over the self-gravitational properties of quantum material systems. In this tradition, the goal is ever to make gravity, so far as concerns quantum material systems in the lab, itself increasingly expansively quantum. 

Our language here is suggestive of \citeauthor{hacking1992self}'s \citeyearpar{hacking1984experimentation, hacking1992self} entity realism, or even (more radically) \citeauthor{latour1987science}'s \citeyearpar{latour1987science} conception of science ``in the making'' read in an ontological mode. Of course, we don't fully embrace Hacking's view, since we take the experiments to be potentially evidential for QG, and not just the systems `sprayed' in the very operation of the apparatus. (And we don't fully embrace Latour's view, since we think there is knowledge to be had.) But like each of them, our emphasis is on achievement of know-how, or novel means of engaging with fundamental physics as a matter of researchers' practice, and specifically in a laboratory setting through managing to isolate delicate systems from environmental interactions. And indeed, the hypothesized outcomes of the newly proposed experiments would confirm extraordinary success in isolating unusually delicate systems of a kind and in a way that we so far do not know how --- something very exciting to anticipate! 

There is another view in the literature against which we should situate ours. \cite{Wallace} describes empirical applications of LEQG (the EFT approach to GR) to various phenomena; since these fall in different physical (sub)regimes, they amount to confirmations of LEQG in each regime, and thence of LEQG in full (up to either the Planck scale or a Landau point). He describes these successful applications as theory `confirmation', and indeed they are from a strict hypothetico-deductive point of view. However, the majority of the applications are in regimes in which very little is at stake: there is little chance of refutation, or of a crucial experiment between LEQG and some alternative. So in a broader and more realistic epistemology, talk of confirmation is overselling the point. (To be fair, Wallace's primary intent is that LEQG actually makes empirical predictions, and the claim about confirmation is secondary.) 

One way of illustrating our criticism is by noting that almost all of the applications he discusses fall in the semi-classical regime, in which the empirical risk is very low. Carrying out the GIE experiments would be an improvement over this situation; if the experiment is successful, and the hypothesized entanglement is observed, then it is incompatible with SCG, and a competitor is ruled out. (Of course, SCG is not generally accepted as a serious competitor, so it ruling out in a crucial experiment, does not provide very strong confirmation.) We note that one of his examples is confirmatory in the same way\footnote{Wallace argues that the Page and Geilker experiment also tests LEQG outside the SCG regime, though we have seen that claim is more controversial.}: the explanation of observed cosmic microwave background structure in terms of fluctuations in the inflaton field in the early universe requires the gravitational field of the early universe to be quantum. (Though the same caveats we mentioned regarding this experiment as a potential witness in footnote \ref{fnCMB} apply to this point too, weakening the claimed confirmation.)

What we would now like to suggest is that from the regimes point of view, the value of performing the GIE experiments is plausibly better captured from the tradition of controlling QG, rather than witnessing it. That is, first: it is compatible with the control tradition that a theorist may help themselves, in their analysis of the experiments, to the interpretive package provided by their favored theories, e.g. LEQG. The control tradition asks for control in ever new regimes, but this demand presupposes a theoretical framework in which regimes can be distinguished; it is the very embrace of LEQG that provides the principled means of distinguishing the physical regimes, in each of which we strive to achieve increased control. And so, taxonomies of noteworthy regimes like that provided by \cite{Wallace} are exactly what is needed for assessing the extent of our control over the quantum nature of gravity \emph{given LEQG}, and for identifying new regimes in which we should seek control --- such as the `coherent-perturbative non-relativistic' regime, probed by GIE.

In the witness tradition, it is trickier to see how distinguishing regimes helps us to understand the value of different experiments. After all, (for a given paradigm) either we have witnessed the quantum nature of gravity in an experiment or we have not. If one succeeded in the first place, why would it be useful to go on and witness QG in another way? Of course, as we have been arguing, there would be all kinds of reasons to perform a witness experiment in another regime. But we see no added epistemic value simply in having witnessed \emph{again} (albeit differently). In other words, we claim that, in the framework set out by Wallace, one can best understand the point of such experiments as GIE from within the control, not witness, tradition. 

The significance of regimes with respect to witnessing is the possibility that different paradigms could set the bar for what counted as observing the quantum nature of gravity in different regimes. For instance, one way to understand the difference between the tripartite and Newtonian paradigms is that the former says that one can witness the quantum nature of gravity in the non-relativistic regime, while the latter requires the relativistic regime. (But observe: this is a point about how to conceptualize the witnessing debate, not a point about the value of achieving a witness experiment.)

Returning to the main thread of this subsection, the upshot of our discussion is that there is a second punchline of the new experiments, if successful: they would affirm that we are able to \emph{make gravity interestingly quantum on the tabletop} in a manner that exceeds our capabilities to do so thus far. And this is something that one can accept regardless of one's paradigm, so regardless of one's view on witnessing. We thus see an unequivocal sense in which the experiment stands to teach us something important and new.

But why, then, is this not the standard line? Why, that is, do we see the primary actors involved as participating rather in the witness tradition, and where, their having embraced the tripartite paradigm, the testing-in-different-regimes benefit of the experiments is rendered merely a satisfying afterthought? We submit the following explanation. Recall the three views of SCG. It strikes us that the witness tradition is very natural (though, we stress, not logically required) on view 1, contrary to view 2: there is an explicit candidate theory standing against taking SCG as fundamental. Refuting view 2, with view 1 the alternative, would seem just to amount to a witness of gravity's being quantum, according to the latter. Meanwhile, the control tradition is very natural on view 3, contrary to view 2: where, to the extent that there is some such micro-physics giving rise to SCG, we should like to learn about it by whatever means avail themselves. But view 3 is just not that popular compared to view 1 --- absent strong explicit rivals to LEQG gathered under the same net, view 3 seems more like a pedantic reminder of the vastness of logical possibility space (so far as concerns means of saving the phenomena). Hence, we offer a genealogical explanation of the present circumstance: in contemplating bringing together quantum theory and gravity, SCG was offered. But it was appealing to understand SCG itself, if/since it could not be fundamental (view 2), specifically in terms of LEQG (view 1).  So experiments pursued were pursued in an effort to put forth increasingly sophisticated witnesses of the quantum nature of that specific underlying theory.

\subsection{Control apart from entanglement: a  `GING' experiment}

A very recent proposal by \cite{Howl} allows us to demonstrate further the importance of recognizing the control tradition: arguments focused too narrowly on a question of QG witness will miss otherwise obvious virtues of novel experimental protocols. They present their tabletop quantum gravity experiment in terms of a witness punchline akin to that for the GIE experiments, but without entanglement as its central mechanism. They presuppose something like the tripartite paradigm for the GIE experiments, now taking a Bose-Einstein condensate's (BEC) transition from a Gaussian to a non-Gaussian state in the presence of a gravitational interaction as indicating that the gravitational mediator is quantum in nature. At its heart, the proposal is based on the insights that (1) non-Gaussianity is measurable in a BEC, that (2) a quantum interaction between the BEC atoms is needed to bring about an overall non-Gaussian state, and that (3) a BEC can be controlled in such a way that no other interaction but the gravitational one effectively occurs between the BEC atoms. The upshot is a Gravitationally Induced Non-Gausianity (`GING') experiment, in parallel with the program of GIE experiments. 

Notably, as we will see below, the experiment yet again cannot make a difference in the in-principle quarrel between the Newtonian model paradigm and the tripartite models paradigm: as in the case of our analysis of the GIE experiments, whether the GING experiment allows for actually testing the quantum nature of gravity is contingent on choices on which reasonable people may disagree. In fact, the analysis of gravity in the GING experiment is, as will be made clear, very similar to the analysis of gravity in the GIE experiments; so it might seem, from that point of view, little advance over the latter. Why not concentrate our efforts, then, on just one? One immediate response comes from a practical perspective: the more cutting-edge scenarios that are available in which we are close to witnessing QG (in the tripartite paradigm), the more likely it is that there is a technical break-through that would make possible our witnessing QG. Another response is more subtle: note that the GING experiment puts us in a different experimental regime. In the context of LEQG for instance, the experiment's use of BECs would seem to probe in a different regime of that theory, compared to the GIE experiments. So, in addition to the merits of testing LEQG in each regime for possible surprises (as, for instance, \cite{Wallace} advocates), perhaps one is interested in regime-specific witnesses of QG: witnessing the quantum nature of gravity \emph{within each notable regime}. However, such a response depends delicately on the way we define regimes: for instance, both experiments fall in the same regime according to Wallace's scheme (see his Table 2): perturbative-coherent-non-relativistic. 

Particularly because of this ambiguity, the latter proposal of seeking \emph{witnesses} across regimes strikes us as idiosyncratic: without careful inspection of one's taxonomy of regimes, it is unclear what virtues there are to achieving a witness of the underlying quantum nature of gravity within each regime, so individuated. One either has witnessed QG or has not (relative to a choice of paradigm) --- no matter the regime in which one has found it. In contrast, per Wallace, there are merits of \emph{testing} across different regimes. And these merits, we claim, are most easily noted in the control tradition, rather than in that focused on witnesses. In the control tradition, every further experimental intervention in every further regime is an advance. This is what we will demonstrate, with the GING experiment as a case study.

To begin, let us supply more details of the experiment, following \cite{Howl}: Gaussian states can be defined as Gaussian-shaped Wigner functions\footnote{The unique Wigner function corresponding to a density state $\hat{\rho}$ for classical position $x$ and momentum $p$ is defined as $W_{\hat{\rho}} (x, p) = \frac{1}{2 \pi} \int dy \exp(-iyp) \braket{x+y | \hat{\rho} | x-y}$.}. Notably though, it is non-Gaussian states that are a key resource of continuous variable quantum information systems, featuring in signature quantum phenomena such as pure state quantum computation or Bell-type experiments.\footnote{See for instance \cite{QuantumComputation1, QuantumComputation2} or \cite{Bell0, Bell1, Bell2}.} (Conversely, Gaussian quantum states may even be effectively simulated using classical systems.\footnote{See, for instance, \cite{ClassicalSimulation}.})

Now, the proposed experimental set-up by \cite{Howl} rests on a Bose-Einstein gas at low temperatures, which can be modeled as a non-relativistic scalar quantum field of form $\hat{\Psi}(r) =\psi(r) \hat{a}$ where $\psi(r)$ is the wave-function of a condensed atom, and $\hat{a}$ the annihilation operator for a condensed atom. Importantly, the gas as such is in a (collective) Gaussian state, with each individual atom in a Gaussian state: the gas is a BEC. More generally, though, the atoms remain in a Gaussian state only as long as they are subject to Gaussian dynamics, that is, to interactions of form \begin{equation} \hat{H} = \sum_k \lambda_k(t) \hat{x_k} + \sum_{k, l} \hat{x_k}^T \mu_{kl}(t) \hat{x_l} \end{equation} where $\hat{x}_k^T := (\hat{x}_k, \hat{p}_k)$, and $\lambda_k(t)$ are 2 x 1 and 2 x 2 real-valued matrices of arbitrary functions of time (a discrete mode spectrum has only been chosen for simplicity).\footnote{For instance, the free scalar field theory with its Hamiltonian
\begin{equation}
\hat{H} = \frac{1}{2} \int d^3  \left[ (\partial_t \hat{\phi})^2 + (\nabla \hat{\phi})^2 + m^2 \hat{\phi}^2\right]
\end{equation} is demonstrably a Gaussian dynamics as can (also) be seen from re-expressing the Hamiltonian in its field modes as $\hat{\phi} = \sum_k u_k(t) \hat{a}_k +v(t) \hat{a}_k^+.$ Quantum Gaussian dynamics essentially act as Bogoliubov transformations on a Gaussian state, that is, Gaussian states are mapped into Gaussian states again.}

Enter Newtonian gravitation then: Newtonian gravitational interactions within the condensate are schematically of form $\int d^3r \hat{\rho(r)} \Phi(r)$ with mass density $\hat{\rho} = m \hat{\Psi}^+(r) \hat{\Psi}(r)$.
Depending on whether the interaction potential $\Phi(r)$ is quantum or classical, one has:
\begin{eqnarray}
\hat{H}_{QG} = \frac{1}{2} m \int d^3 r :\hat{\Psi}^+(r) \hat{\Psi}(r) \hat{\Phi}(r):\\ 
\hat{H}_{CG} = m \int d^3 r \hat{\Psi}^+(r)  \hat{\Psi}(r) \Phi[\Psi]] (t, r)
\end{eqnarray}
where : denotes normal ordering. Using the fact that $\hat{\Psi}(r) \approx \psi(r) \hat{a}$ where $\hat{a}$ is the annihilation operator for the condensate and $\psi(r)$ the wavefunction for the individual atom, we see that the interaction Hamiltonian is non-Gaussian for the quantum interaction and Gaussian for the classical interaction

\begin{eqnarray}
\hat{H}_{QG} = \frac{1}{2}\lambda_{QG} \hat{a}^+ \hat{a}^+ \hat{a} \hat{a}\\
\hat{H}_{CG} = \lambda_{CG}[\Psi] \hat{a}^+ \hat{a}
\end{eqnarray}
with 

\begin{eqnarray}
\lambda_{QG} := - Gm^2 \int d^3 r' \frac{ |\psi(r')|^2 |\psi(r)|^2}{|r-r'|}\\
\lambda_{CG}[\Psi] := Gm \int d^3 r |\psi(r)|^2 \Phi[\Psi] (t, r).
\end{eqnarray}

In other words, if (i) the system becomes non-Gaussian and (ii) gravity is the only interaction at play, then we have a witness of the quantum nature of gravity, or so the argument goes. Enticingly, it is expected that within the near future it will indeed be technologically possible to screen off all other but the mutual gravitational interactions. More concretely, at low energies the BEC will feature Van der Waals interactions, as well as a gravitational interaction. But by applying a suitable tuned external magnetic field, one can reduce the s-wave scattering length in the BEC to zero and thus switch off the Van der Waals interactions (for details, see \cite{Howl}). Needless to say, the universal gravitational coupling cannot be screened off. 

Interestingly, the quantitative measure of how much the effect of gravitational interaction dominates other interactions --- the signal-to-noise ratio for gravitational-induced non-Gaussianity of a single measurement on the BEC --- can be put into analogous form to the phase in the GIE experiment. To see this, observe that if $M$ is the mass of the BEC, $\delta \tau := \sqrt{2/\pi} G Mt/Rc^2$ with $t$ the interaction time, $M_P$ the Planck mass, and $t_p$ the Planck time, then:

\begin{equation}
    \frac{M}{M_P} \frac{\delta \tau}{t_P} = \frac{ct}{R}\left(\frac{m}{M_P}\right)^2.
\end{equation}
Compare this expression to the phase in equation (6) for the GIE experiment, which, after restoration of units, becomes 

\begin{equation}
    \frac{G m^2 t}{\delta} = \frac{ct}{\delta} \left(\frac{m}{M_P}\right)^2.
\end{equation}
The only difference between the equations lies in the characteristic length scales of the respective set-ups, that is, between whether the formula features the size of the BEC $R$ or the distance of separation $\delta$.

But where does this new proposal leave us? According to its proponents, it has various possible advantages and disadvantages over the GIE-experiment: (1) The set-up concerns a `single' matter system, whereas the GIE-experiment features two matter systems. But this is only an advantage if the point is intended in a practical manner: from a more ontological point of view, what we are probing of the BEC are rather interactions of millions of condensed atoms. (2) In step 1 of our proof that gravitationally induced entanglement cannot be explained by a classical mediator, we make the assumption (a) that the gravcats interact locally (i.e., that they have no mutual interaction term in the Hamiltonian). Without that assumption, the proof fails (for instance, if one supposes some non-local physics that supplements the standard account of gravity in the Newtonian limit). In the GING experiment no such assumption is required, and so this would seem an advantage.(3) A disadvantage is that non-Gaussianity (as opposed to GIE-type entanglement) can be created by collapse-type scenarios. (4) The description has not been given a more abstract formulation as that of the GIE-type experiments yet; the description proceeds under recourse to straightforward QFT, which could be unattractive. (5) \cite{Howl} claim that their experiment only really requires ``the simple process of adding a `hat' to classical gravitational degrees of freedom" (p. 010325-2) to witness QG --- GIE by contrast involves background discussions about locality and the significance of gauge constraints. But such a claim would entirely undermine the appeal of the experiment as capable of witnessing the quantum nature of gravity! Namely: if all that was required was an experiment which required us to add a `hat' to the Newtonian potential, then this was already done in the 1970s by COW. 

So, advantages and disadvantages --- just so, they're different experiments, different kinds of systems! In the witness tradition, however, the virtues of performing both seem easily overlooked. In particular, it still does not settle the quarrel between would-be proponents of the Newtonian model and tripartite models paradigms. The proponents of the tripartite models paradigm would seem equally excited about a claim of witness in this new experiment, while the Newtonian model paradigm proponents would once more insist that the experiment fails to provide a witness, precisely because of the vindication of the naive Newtonian description of the experiment. All told, the hopes of the experiments in the witness tradition would seem to rise and fall together; the virtues of diversity of experiment are lost. By contrast, in the control tradition the mixed bag of `advantages' and `disadvantages', claimed or otherwise, of the GING experiment over the GIE experiment is reinterpreted as an explicit discussion of the differences in the kind of access to or control of the gravitational interaction in various quantum experiments.

\section{Concluding remarks: taking stock of quantum gravity phenomenology}

According to lore in the philosophy of QG, the problem of quantum gravity is (very nearly) purely one for the theoreticians. It is just too difficult to hope for discriminating signatures of QG in data, because the relevant empirical regimes far exceed our capacities for experimentation (in high energy physics) or direct detection (in astrophysics and cosmology). But this lore is misleading of fundamental physics practice today. In recent decades, and to wide acclaim in the surrounding discipline, a range of empirical testing strategies have been pursued within the arena of quantum gravity phenomenology, as proposed means of gaining significant, increased empirical traction on the problem of QG.   

Our focus throughout this manuscript has been on disentangling threads in the interpretation of experiments within one such emerging empirical testing strategy: tabletop quantum gravity. Our impetus was evident disagreement among physicists involved in tabletop quantum gravity concerning a new class of proposed experiments that are otherwise by and large constitutive of the new subject.\footnote{In fact, there is, arguably, another class of experiments in tabletop quantum gravity that we ignored: experiments attempting to demonstrate indefinite causal order on the tabletop, e.g. \citep{rubino2017experimental}. Now,there is a sense in which conformal structure is a more profound component of the spacetime metric in GR, and so perhaps experiments focused on producing superposition states of conformal structure ought to be distinguished from the rest. It seems dubious to us that there are, indeed, ultimately satisfying reasons to treat conformal structure in this privileged way, however. But even so, we hope that the careful attention we have given to the question of the relationship between QG witness and metric superposition states by means of GIE experiments is informative in thinking through possible upshots of these other proposed experiments.} Hopefully, we have clarified that such disagreement reflects something more subtle and foundationally noteworthy than a simple confusion by some parties to the dispute. On one hand, the Newtonian model counseled against any claim that the GIE experiments would, by hypothesis, provide a witness of the quantum nature of gravity, given the interpretational framework provided by the quantization of gauge theories. On the other hand, the tripartite models counseled in favor of the very same claims, given treatments of gravity as a mediating subsystem. Thus, one has the following moral: how we assess such experiments is a much more intricate affair than may be at first thought, and crucially depends on matters of physical interpretation. 

This is perhaps not a surprising statement in the general philosophy of experimentation in physics (or more generally!). Disagreements over the stakes of experiments or observation are almost certainly to be found in disagreements over interpretations of the naive descriptions assigned to the relevant experimental set-ups, given the sum total of the lessons provided about the world by our current best physics. And just so: our explanation of the disagreement over the claim of quantum gravity witness in the new GIE experiments in tabletop quantum gravity was sensitive to paradigmatic disagreement about the experiments, understood by the lights of our current best physics. Still, it is interesting to see how the general thesis shows up, in particular, given the specific details of contemporary QG research: where most theorists involved already endorse LEQG en route to developing a future high-energy theory of QG, but meanwhile there are more agnostic approaches available to merging together our current corpus of physics, in anticipation of that future theory. And so, there is a shifting sense within quantum gravity phenomenology as to what properly amounts to `our current best physics', to be brought to bear on the analysis of any laboratory gravitational experiment with quantum matter probes. And from this perspective, the new class of experiments is arguably a first arena where that shiftiness spells fierce disagreement among those otherwise allied in an empirical-first pursuit of a high-energy theory of quantum gravity.

Nevertheless, as we saw in the previous section, there are means of packaging the empirical stakes of such tabletop quantum gravity experiments that are largely immune to possible foci of disagreement over the matters of interpretation. Great care must therefore be given to whether bottom-line endorsements of new experiments on the horizon are defended on argumentative grounds sensitive to interpretation (in the sense intended here), or on argumentative grounds that are separable from such interpretive commitments. In particular, the tradition of discussing experiments in tabletop quantum gravity in terms of witnesses of the quantum nature of gravity is rife with troubles born by such sensitivity, while the tradition that emphasizes control over the nature of the (typically presumed quantum) gravitational coupling between quantum matter probes seems more promising for cashing out the stakes of such experiments. In certain contexts, this tradition would seem to reduce to talk about testing in various regimes, but we have argued that such a position is derivative of further (common) assumptions about the relevance of LEQG, which need not be embraced to appreciate the same experiments. 

From all of this, two metaphilosophical lessons on future work immediately follow: First, philosophy of physics (especially philosophy of QG) usually focuses on theory and is little interested in experiments. We take our work to show that in times of crisis, clarifications of what we mean and intend with certain experiments can be just as philosophically exciting (and important) as what we normally discuss with relation to pure theory. Second, with our book, we hope to have provided a case for a philosophy \textit{in} physics (in the style of what \cite{pradeu2021philosophy} call ``philosophy in science") that engages with an on-going debate in physics in time and not from hindsight, and which operates (at least) with the (side) goal of being of actual use to the practitioner. In fact, what one could witness, or so we hope, is a prime example of how an outstanding controversy between physicists can at times \textit{only} be settled by making recourse to the philosophers' toolbox. It is in these two respects that we take philosophers of physics can contribute more to the search of a theory of quantum gravity!

\bibliographystyle{plainnat}
\bibliography{references.bib}

\newpage
\section{Appendix A: Mannheim's analysis of the COW experiment}

\begin{figure}[htbp]
\begin{center}

\begin{tikzpicture}[scale=1]

\draw[ultra thick] (-.75,-.75) -- (.75,.75); 
\draw[ultra thick] (3,3) -- (4.5,4.5);
\draw[ultra thick] (-.5,3.5) -- (.5,4.5); 
\draw[ultra thick] (3.5,-.5) -- (4.5,.5); 
\draw[|<->|] (5.75,0) -- (5.75,2) node[anchor=west] {h} -- (5.75,4);
\draw[|<->|] (0,-1) -- (2,-1) node[anchor=north] {h} -- (4,-1);

\draw[thick,->] (-1,0) node[anchor=south east] {S} -- (0,0) node[anchor=north west] {A};

\draw[thick,dashed,->] (0,0) -- (4,0);
\draw[thick,->] (0,0) -- (3.75,-.25) node[anchor=north west] {B$_1$};

\draw[thick,dashed,->] (4,0) -- (4,4);
\draw[thick,<-] (3.25,3.25) node[anchor= east] {D$_2$} -- (3.75,-.25);

\draw[thick,->] (0,0) -- (0,4) node[anchor=south east] {C};

\draw[thick,dashed,->] (0,4) -- (4,4);
\draw[thick,->] (0,4) -- (3.75,3.75) node[anchor= north] {\ D$_1$};

\draw[|<->|] (4.5,3.75) -- (4.5,3.88) node[anchor=west] {$\delta$} -- (4.5,4);
\draw[|<->|] (4.5,0) -- (4.5,-.13) node[anchor=west] {$\delta$} -- (4.5,-.25);
\draw[|<->|] (3.75,4.5) -- (3.88,4.5) node[anchor=south] {$\delta$} -- (4,4.5);

\draw[|<->|] (5,3.25) -- (5,3.63) node[anchor=west] {$3\delta$} -- (5,4);
\draw[|<->|] (3.25,5.25) -- (3.63,5.25) node[anchor=south] {$3\delta$} -- (4,5.25);

\end{tikzpicture}
\caption{The actual paths followed by a neutron along the different paths, taking into account the effect of gravity on the classical motion (and approximating parabolic motion with straight paths).}
\label{fig:COWMann1}
\end{center}
\end{figure}

In this appendix, we sketch the highlights of Mannheim's treatment. To analyze the path, we set up coordinates with A at the origin, and $x$ and $y$ the horizontal and vertical axes, respectively. Referring to figure 
\ref{fig:COWMann1}, consider the lower path first: instead of the horizontal path AB, a neutron will follow a parabola, $y=-gx^2/2u^2$, until it intersects the mirror lying along $y=x-h$. To first order in $g$ this is the point $(h-\delta,-\delta)$, where $\delta=gh^2/2u^2$, the exact distance that the neutron would have fallen had it reached $x=h$. Again to first order in $g$, the distance traveled by the neutron is $h-\delta$, and its average speed is $u$: hence the phase change along AB$_1$, $\Delta\phi_{AB_1}$, is given by $pr = mu(h-\delta)$. The neutron is not deflected from the path AC by gravity, though it decelerates from an initial velocity of $u$ to a final velocity of $u-gh/u$ to first order in $g$, as previously noted. Thus its average velocity is $u-gh/2u$, and  $\Delta\phi_{AC}=m(u-gh/2u)h=mu(h-gh^2/u^2)=mu(h-\delta)$. Thus $\Delta\phi_{AB_1}=\Delta\phi_{AC}$, the phase changes along the two initial path segments are equal! 

The calculation of the second two path segments continues in the same way. The component leaving C does not reach D, but by similar considerations to the AB$_1$ path reaches, to first order, a point D$_1$ with coordinates $(h-\delta,h-\delta)$, traveling a distance $h-\delta$ at an average speed $u-gh/u$; thus $\Delta\phi_{CD_1}=mu(h-3\delta$ plus terms higher order in $g$. The considerations for the final path segment are that (a) it starts at $(h-\delta,-\delta)$, (b) it is not reflected into the horizontal but with a velocity $(-ghu,u)$, and (c) suffers vertical  deceleration. We leave the calculation to the reader, but heuristically, in traveling to the line $x=y$ (at which the waves recombine), by (b) the neutron travels $2\delta$ horizontally; since by (a) it started at $x=h-\delta$, it arrives at a point D$_2$ with coordinates $(h-3\delta,h-3\delta)$. To first order, the length of B$_1$D$_2$ is given by $h-2\delta$, and by (c) the average velocity is $(u-gh/2u)$, so that $\Delta\phi_{B_1D_2}=m(u-gh/2u)(h-2\delta)=mu(h-3\delta)=\Delta\phi_{CD_1}$ -- the phase changes along the second pair of paths are also equal! (And of course, $\Delta\phi_{AC}\neq\Delta\phi_{B_1D_2}$, even though $\Delta\phi_{AC}=\Delta\phi_{BD}$.) 

\begin{figure}[htbp]
\begin{center}

\begin{tikzpicture}[scale=1]

\draw[ultra thick] (-.75,-.75) -- (.75,.75); 
\draw[ultra thick] (3,3) -- (4.5,4.5);
\draw[ultra thick] (-.5,3.5) -- (.5,4.5); 
\draw[ultra thick] (3.5,-.5) -- (4.5,.5); 


\draw[thick,->] (-1,.5) -- (.5,0.5) node[anchor=north west] {A$_1$};
\draw[thick,->] (-1,0) node[anchor=south east] {S} -- (0,0) node[anchor=north west] {A};

\draw[|<->|] (0,.75) -- (.25,.75) node[anchor=south] {$2\delta$} -- (.5,.75);
\draw[|<->|] (3,3.25) -- (3,3.5) node[anchor=east] {$2\delta$} -- (3,3.75);
\draw[|<->|] (3.25,4) -- (3.5,4) node[anchor=south] {$2\delta$} -- (3.75,4);
\draw[|<->|] (0,0) -- (0,.25) node[anchor=east] {$2\delta$} -- (0,.5);

\draw[thick,dashed,->] (0,0) -- (3.75,-.25) node[anchor=north west] {B$_1$};
\draw[thick,->] (0.5,.5) -- (4.25,.25) node[anchor=north west] {B$_2$};
\draw[gray,ultra thick,->] (1.875,-.125) -- (2.375,.375);

\draw[thick,dashed,<-] (3.25,3.25) node[anchor= west] {D$_2$} -- (3.75,-.25);
\draw[thick,->] (4.25,.25) -- (3.75,3.75);
\draw[gray,ultra thick,->] (3.5,1.5) -- (4,2);

\draw[thick,->] (0,0) -- (0,4) node[anchor=south east] {C};

\draw[thick,->] (0,4) -- (3.75,3.75) node[anchor= west] {D$_1$};



\end{tikzpicture}
\caption{The actual paths followed by a (point on the wavefront of a) neutron along the different paths, taking into account the effect of gravity on the classical motion (and approximating parabolic motion with straight paths). The phase changes along A$_1$B$_2$D$_1$ and ACD$_1$ have been shown to agree, but SA$_1$ is $2\delta$ longer than SA, accounting for the observed relative phase.}
\label{fig:COWMann2}
\end{center}
\end{figure}
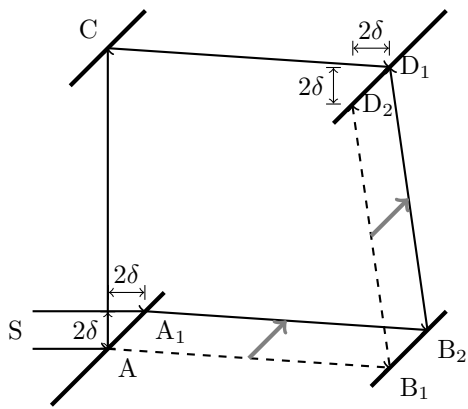

But since $\Delta\phi_{AB_1}=\Delta\phi_{AC}$ and $\Delta\phi_{B_1D_2}=\Delta\phi_{CD_1}$ we have that $\Delta\phi_{AD_2}=\Delta\phi_{AD_1}$ -- there is no relative phase shift around the closed path! What then is the source of the observed interference? Or is there a flaw in Mannheim's calculation? The resolution rests on the fact that the two paths do not meet at a point, and so cannot describe interference: it makes no sense to add the phases at D$_1$ and D$_2$! So we take into account that a neutron beam does not follow a 1-dimensional path, but has area, and so occupies a volume; so the paths we have been considering track only a single point of the beam cross-section through the apparatus. To compute the relative phase we need to consider the phase shift of a point that follows the lower path, not to D$_2$, but to D$_1$; in other words, that follows a path offset by $(2\delta,2\delta)$ from AB$_1$D$_2$, namely A$_1$B$_2$D$_1$ shown in figure \ref{fig:COWMann2}. The shift clearly leaves the phase shift along the two legs after splitting unchanged, and equal to the two legs after splitting along the path via C. Equally (assuming that the wave at the source is coherent across its area, and indeed the time it takes the neutron to traverse the distance), the shift clear introduces an extra $2\delta$ into the path between source and splitter, leading to an extra phase of $2mu\delta=mgh^2/u$ in the lower versus upper path -- the very effect observed by COW. To repeat: in Mannheim's account, the entire effect is due to the extra distance travelled to reach the beam splitter, not from the path around the apparatus, as OW assumed.

\section{Appendix B: A Newton-Cartan analysis of gravcats}

Here we derive a general expression for differential contributions to the phase factor for a quantum test particle of mass $m$ subject to Newtonian gravity, due locally to the Newton-Cartan spacetime geometry in the neighborhood of the particle. (`Differential' in the sense that the phase is defined relative to a choice of zero-point: the spacetime geometry for which there is understood to be no such additional contribution to the phase.) The following is intended to be analogous to the analysis provided in the (linear, post-Newtonian regime) general relativistic case by \cite{christodoulou2019possibility}, discussed in \S\ref{sec:Bose}.

As in their analysis, the basis states of the gravitational field in a gravcat QG witness experiment are assumed to correspond, in the neighborhood of either one of the gravcats, to what may be described classically as the local gravitational field. This  field is associated with a self-gravitation effect, as well as an effect due to a source mass (the other gravcat) at some spatial distance that remains approximately fixed (for the short lifetime of the experiment). Following \cite{christodoulou2019possibility}, we assume that in all but one classical configuration of the two gravcats ---  the one in which they are closest together --- the gravitational self-contribution of any one particle to its phase factor is so great as to dominate the effect of the source particle, at least over the lifetime of the experiment. This self-contribution thus defines the zero-point phase contribution, and we only need determine how the state in which the gravcats are closest modifies it to find the differential contribution.\footnote{That is, the phase contribution is effectively given (up to a multiple) by the total energy \textit{over time} associated to the two gravcats closest together. Cf. \cite{christodoulou2019possibility}.}

Formally, let $(M,t_a,h^{ab},\mathring{\nabla})$ be a classical spacetime: $t_ah^{ab}=\textbf{0}$ for all $p\in M$ and $\mathring{\nabla}$ is a flat derivative operator on $M$ compatible with $t_a$ and $h^{ab}$. For some curve (with endpoints) $\gamma:[s_1,s_2]\rightarrow M$ modeling the worldline of one of the two gravcats --- idealized as a test particle of mass $m$ located at its own center of mass --- through the duration of the experiment, we will assume that $(M,t_a,h^{ab},\mathring{\nabla})$ is apt as a model of the geometric features of the experimental setup, at least in a sufficiently small neighborhood of $\gamma$ (note that which of the two gravcats we pick for this purpose does not matter, due to the symmetries of the experimental setup). Consider two Newton-Cartan models of Newtonian gravity in the neighborhood of $\gamma$, defined respectively as $(M,t_a,h^{ab},\nabla)$ and $(M,t_a,h^{ab},\nabla')$: that is, the models differ at most in their curved derivative operators $\nabla$ and $\nabla'$. Without loss of generality, we may regard these two models as Trautman geometrizations of two models of standard Newtonian gravitation on $(M,t_a,h^{ab},\mathring{\nabla})$, corresponding to gravitational potentials $\phi$ and $\phi'$ (satisfying Poisson's equation on $(M,t_a,h^{ab},\mathring{\nabla})$). In the neighborhood of $\gamma$, $\phi$ and $\phi'$ may, as noted, be decomposed into a self-gravity term (identical in each case) and a term due to the gravcat's interaction with the other gravcat situated at a spatial distance, fixed for the duration of the experiment. In what follows we will take $\phi'>\phi$ so that $(M,t_a,h^{ab},\nabla')$ is the geometrized model of the gravitational field when the other gravcat is nearest. For convenience then, we will speak of $\nabla'$ as the ``errant'' geometry, relative to the standard set by $\nabla$, i.e. the (geometrized) gravitational model $(M,t_a,h^{ab},\nabla$) that constitutes the ``lab-average''. Assume from hereon that $\gamma$ is geodesic with respect to $\nabla$, but whether the gravcat traversing $\gamma$ is understood to survey $\nabla$ or the errant geometry $\nabla'$ depends on the configuration of the two gravcats. Since their separation  will differ across branches of the joint wavefunction (\ref{eq:BSspin}), one immediately sees that there will be branch-relative differences in spacetime geometry, reflecting branch-relative differences in the local gravitational field about $\gamma$. 

From the Trautman geometrization lemma, we know that $\nabla=(\mathring{\nabla},-t_at_bh^{cd}\nabla_d \phi)$ and $\nabla'=(\mathring{\nabla},-t_at_bh^{cd}\nabla_d \phi')$,\footnote{This notation is explained in \cite{malament2012topics}.} where in each case we exploit the fact that $\mathring{\nabla}_a \psi=\nabla_a \psi = \nabla'_a \psi$ for any scalar field $\psi$. From this, one can easily show that, since $\xi^m\nabla_m\xi^a=\textbf{0}$ for the four-velocity $\xi^a$ of the gravcat (that is, whose integral curve is $\gamma$), then $\xi^m\nabla'_m\xi^a= \xi^m\nabla'_m\xi^a - \xi^m\nabla_m\xi^a=\xi^m\xi^nt_mt_nh^{ab}\nabla_b(\phi'-\phi)$. In other words, the acceleration of $\xi^a$ with respect to $\nabla'$ relevant in the one branch of the wavefunction, given that the integral curve $\gamma$ is geodesic with respect to $\nabla$, is fully determined by the difference in potentials between the one branch and the rest. (Note that this expression is fully general: if $\phi=\phi'$ up to the addition of a constant, then $\nabla=\nabla'$ up to a constant multiple, and so the acceleration of the gravcat that surveys the errant geometry can be seen to vanish necessarily with respect to $\nabla'$, given that the integral curve is geodesic with respect to $\nabla$.)

We may now consider the quantity of power that must be input into the gravcat at any point in its trajectory --- in the branch in which it surveys the errant geometry $\nabla'$ --- relative to the zero power that is needed for the same trajectory in the branches whose gravitational field approximates the lab average spacetime geometry $(M,t_a,h^{ab},\nabla)$. Since acceleration is spacelike, we know that there exists\footnote{See proposition 4.1.1 and the discussion immediately preceding it in \cite{malament2012topics}.} a covector $u_b$ such that $h^{ab}u_b=\xi^m\xi^nt_mt_nh^{ab}\nabla_b(\phi'-\phi)$. So, for $Power=(m \cdot u_b)\xi^b=m \cdot (u_b\xi^b)$ (`power = force $\times$ velocity'), it follows that

\begin{equation}
Power=m\xi^b (\xi^m\xi^nt_mt_n)\nabla_b(\phi'-\phi).
\end{equation}

\noindent It is important to stress that this quantity of power is defined pointwise along $\gamma$. What this means is that we may consider the integral of this expression over the curve $\gamma$ to compute the total power associated with the gravcat's surveying $\nabla'$, through the course of its geodesic path in the lab-average spacetime.

Finally, we may consider the amount of (virtual) work done on the gravcat surveying $\nabla'$ at each instant along $\gamma$: the quantity of power multiplied by the `proper'\footnote{Here, `proper' just notes that the scalar quantity is again defined pointwise, i.e. as a degenerate tensor; it is a further fact about classical spacetimes that, integrated over a curve between endpoints, proper time elapsed along the curve necessarily agrees with the global time elapsed between the two endpoints.} time $t_c\xi^c$ experienced by the particle at that instant.\footnote{As observed by \cite{weatherall2014geometry}, we may, in effect, regard this quantity of work as the result of an ordinary force field on $(M,t_a,h^{ab},\nabla)$, which acts only on particles there that would survey the errant $\nabla'$. Meanwhile, the force field that acts there on particles that survey $\nabla$ is the $\textbf{0}$ tensor, and so if $\nabla'=\nabla$ up to a constant multiple (corresponding to $\phi'=\phi$ up to an additive constant) zero work is put into the particle. Note that, as an upshot, the computed quantity of energy here is no more or less mysterious than that which is put into any charged particle in the presence of a fixed, ambient field, which deflects that charged particle off of its geodesic path in accordance with a force law associated with the field-charge pair. This will be important shortly, in promoting this quantity of total energy to a quantum phase factor.} Integrated over the curve $\gamma$, this scalar product of power and proper time yields an expression for the total energy input to the gravcat over the lifetime of the experiment, with respect to $(M,t_a,h^{ab},\nabla)$, for its surveying the errant $\nabla'$:

\begin{equation}
    Total_{~}Energy = m\int_\gamma{\xi^b(\xi^m\xi^n t_m t_n)\nabla_b(\phi'-\phi) (t_c\xi^c) dS}.
\end{equation}

In the specific case of the experiment, for either choice of gravcat, we may note that --- at least to good approximation over the short lifetime of the experiment --- the difference $\phi'-\phi$ is symmetric under time translations for the duration. In other words, adopting coordinates appropriate for the experimental setup (for definiteness, considered in the lab-average case), this entire expression evaluates to $m(\phi'-\phi)*T$ where $T$ is the total time elapsed and $(\phi'-\phi)$ is understood as a function solely of spatial coordinates at a point. (This follows from separating out the expression for the total power from the expression for the total proper time elapsed. That the second yields $T$ is trivial; that the first yields $m(\phi'-\phi)$ follows from the time translation symmetry of the difference $\phi'-\phi$.) Moreover, noting that the self-gravity contributions to the potential about $\gamma$ cancel, it is easy to see that the differential contribution to the phase factor provided by the errant geometry relevant in the one branch relative to the rest, which is there due to the nearer --- hence, stronger --- mutual gravitational attraction of the two gravcats so geometrized, is identical to that which is calculated in the main text, in what was there dubbed the Newtonian model. 

To see that the expression calculated is the correct quantity, consider the following prescription for quantization, based on the work of \cite{bohm1987ontological} (though we will not be taking their `ontological' interpretation here). They note that if one writes the wavefunction as $\psi=Re^{iS/\hbar}$, with $R=|\psi|$ and $S$ the phase, then the time-dependent Schr\"odinger equation 
\begin{equation}
    \label{eq:TDSE}
    i\hbar\partial_t\psi=-\frac{\hbar^2}{2m}\vec{\nabla}^2\psi+V\psi
\end{equation}
yields for the real and imaginary parts (respectively):

\begin{equation}
    \label{eq:QH-J}
    \partial_tS +(\vec{\nabla} S)^2/2m + V -\hbar^2 \vec{\nabla}^2R/2mR=0,
\end{equation}
and
\begin{equation}
    \label{eq:Bohm2}
    \partial_tR^2+\vec{\nabla}(R\vec{\nabla} S)/m=0.
\end{equation}
Or, in the case that $R$ is a constant with respect to space and time, as for instance in the plane wave that describes a gravcat state, the first reduces to 

\begin{equation}
    \partial_tS +(\vec{\nabla} S)^2/2m + V = 0,
\end{equation}
which is the Hamilton-Jacobi equation for corresponding classical particle (recall that $\vec{p} = \vec{\nabla} S$), with $S$ now interpreted as the action. Or, reversing our reasoning to this point, we can take as an equivalent quantization procedure for such systems that one takes the classical action and promotes it ($\times\ i/\hbar$) to the phase of a wavefunction.\footnote{Now, because we also have (\ref{eq:Bohm2}) we do not have classical particle mechanics: as is well-known, in Bohmian mechanics the wavefunction determines the velocity of a particle at any point, not (only) its acceleration. Nevertheless, Bohm and Hiley propose understanding the last term in (\ref{eq:QH-J}) as a `quantum potential'. All this is of course besides the point in the gravcat system in which $R$ is constant and the term vanishes.} But the quantity that we just computed for the Newton-Cartan gravcats, namely total energy over time, effectively is the classical action (there is no potential term so that the time integral over the Hamiltonian (total energy) and over the Lagrangian (classical action) are effectively the same), and thus precisely is the phase that we seek.\footnote{Similarly, for the Newtonian formulation of the gravcats, Hamiltonian and Lagrangian are the same up to a sign due to the assumption of a static set-up (now there is no kinetic term so that the time integral over the Hamiltonian (total energy) and over the Lagrangian (classical action)).}
\end{document}